\documentclass[%
nofootinbib,
amsmath,amssymb,
aps,
pra,
sectionbib
]{revtex4-2}

\usepackage{graphicx}
\usepackage{dcolumn}
\usepackage{bm}

\usepackage[caption=false]{subfig}
\usepackage[table]{xcolor}
\usepackage{newtxtext,newtxmath}
\usepackage[T1]{fontenc}
\usepackage{tabularx}
\usepackage{cancel}
\usepackage[colorlinks = true,citecolor=blue,linkcolor = blue,urlcolor=blue,breaklinks = true]{hyperref}
\usepackage{cleveref}
\usepackage{multirow}
\usepackage{float}
\usepackage{colortbl}
\usepackage{url}
\usepackage{orcidlink}

\crefformat{figure}{Fig.~#2#1#3}
\crefformat{equation}{equation~#2#1#3}
\crefformat{section}{section~#2#1#3}
\crefformat{table}{Tab.~#2#1#3}
\crefformat{appendix}{appendix~#2#1#3}
\crefmultiformat{figure}{Figs.~#2#1#3}{~and~#2#1#3}%
    {,~#2#1#3}{,~#2#1#3}
\crefrangeformat{figure}{Figs.~(#3#1#4--#5#2#6)}

\definecolor{LightCyan}{rgb}{0.88,1,1}

\definecolor{VirdisA}{HTML}{443983}
\definecolor{VirdisB}{HTML}{31688e}
\definecolor{VirdisC}{HTML}{21918c}
\definecolor{VirdisD}{HTML}{35b779}
\definecolor{VirdisE}{HTML}{90d743}
\definecolor{VirdisF}{HTML}{fde725}

\newif\ifhighlightchanges
\highlightchangesfalse
\newcommand{\change}[1]{{\ifhighlightchanges\color{red} #1\else #1\fi}}

\begin{document}

\title{The mass distribution of the first stars can be determined via the 21-cm signal}

\author{T. Gessey-Jones\,{\orcidlink{0000-0002-4425-8746}}$^{\,1,2}$,
N. S. Sartorio\,\orcidlink{0000-0003-2138-5192}$^{\,3}$,
H. T. J. Bevins\,\orcidlink{0000-0002-4367-3550}$^{\,1,2}$,
A. Fialkov$^{2,4,\change{*}}$, 
W. J. Handley\,\orcidlink{0000-0002-5866-0445}$^{\,1,2}$, \\
E. de Lera Acedo\,\orcidlink{0000-0001-8530-6989}$^{\,1,2}$, 
G. M. Mirouh\,\orcidlink{0000-0003-0238-8435}$^{\,5,6}$,
R. G. Izzard\,\orcidlink{0000-0003-0378-4843}$^{\,7}$,
and R. Barkana\,\orcidlink{0000-0002-1557-693X}$^{\,8}$
\\
$^{1}$Astrophysics Group, Cavendish Laboratory, J. J. Thomson Avenue, Cambridge, CB3 0HE, UK\\
$^{2}$Kavli Institute for Cosmology, Madingley Road, Cambridge, CB3 0HA, UK \\
$^{3}$Sterrenkundig Observatorium, Ghent University, Krijgslaan 281-S9, B-9000 Gent, Belgium\\
$^{4}$Institute of Astronomy, University of Cambridge, Madingley Road, Cambridge, CB3 0HA, UK \\
$^{5}$Departamento de Física Teórica y del Cosmos, Universidad de Granada, Campus de Fuentenueva s/n, 18071 Granada, Spain\\
$^{6}$Instituto de Astrofísica de Andalucía (CSIC), Glorieta de la Astronomía s/n, 18008 Granada, Spain\\
$^{7}$\change{School of Physics and Mathematics}, University of Surrey, Guildford, GU2 7XH, Surrey, UK\\
$^{8}$Department of Astrophysics, School of Physics and Astronomy, Tel-Aviv University, Tel-Aviv, 69978, Israel \\
$^{*}$\change{afialkov@ast.cam.ac.uk}
}

\date{\today}

\begin{abstract} 
\textbf{
The formation of the first stars and the subsequent population of X-ray binaries represents a fundamental transition in the state of the Universe as it evolves from near homogeneity to being abundant in collapsed structures such as galaxies. 
Due to a lack of direct observations, the properties of these stars remain highly uncertain. 
By considering the impact of the first stars and their remnant X-ray binaries on the cosmological 21-cm signal, we demonstrate that upcoming observations \change{have the potential to significantly improve our understanding of these objects.}
We find a 25\,mK sensitivity measurement of the 21-cm global signal by a wide-beam radiometer, such as REACH, or 3,000 hours of foreground avoidance observations of the 21-cm power spectrum by SKA-Low, 
\change{could} provide three-sigma constraints on the mass distribution of the first stars.
Such measurements will fill a critical gap in our understanding of the early Universe and aid in interpreting high-redshift galaxy observations. 
}
\end{abstract}
\maketitle

Observations of the cosmic microwave background (CMB) show us that $\sim 380,000$ years after the Big Bang, the Universe was nearly homogeneous, and visible matter at this time was composed almost entirely of hydrogen and helium. 
In contrast, the Universe seen today is abundant in heavier elements, 
and approximately $10\%$ of visible matter has collapsed into stars and galaxies. 
Building a complete picture of the transition between these disparate states is a principal focus of modern cosmology, 
a vital part of this picture being the first generation of stars and their compact object remnants.

The first generation of stars, the Population III (Pop~III) stars, must have formed at cosmic dawn from the unenriched gas that permeated the Universe post-recombination~\citep{Klessen_2023}.
They will have produced the first heavier elements~\citep{Bromm_2001} and re-illuminated the Universe, thus ending the cosmic dark ages~\citep{Barkana_2016} and beginning the reionization of the intergalactic medium (IGM).
Since these stars are composed of only hydrogen and helium, their formation mechanism, properties, evolution, and ultimate fates are believed to have been distinct from the later generations of stars we observe today~\citep{Klessen_2023}.
However, there remains disagreement about the expected properties of the first stars, including their initial mass function (IMF, the mass distribution of stars when they reach the main sequence).
For example, hydrodynamic simulations of Pop III star formation vary in their predictions for the median stellar mass by $2.5$ orders of magnitude~\citep{Stacy_2013, Hirano_2014, Hirano_2015, Wollenberg_2020, Prole_2022, Jaura_2022}.
This uncertainty in mass distribution, in turn, propagates into uncertainties on the impacts of the first stars, such as their contributions to the metal enrichment~\citep{Pallottini_2014}, reionization~\citep{Schaerer_2002}, and heating~\citep{Sartorio_2023} of the \change{high-redshift} IGM, the latter principally occurring via their remnant X-ray binaries (XRBs) \change{\citep{Fragos_2013}, although cosmic rays may also significantly contribute to heating~\citep{Gessey-Jones_2023}. 
}

Until recently, experimental answers to the uncertain Pop III IMF seemed out of reach, with the first stars too distant and too faint to be directly observed individually~\citep{Schauer_2020}.
However, stellar archaeology provides some insights, with the null-detection of surviving Pop~III stars suggesting first-generation stars with masses $< 0.8$\,M$_{\odot}$~\citep{Magg_2019} were rare, and the recent potential detection of the metal signatures of a pair-instability supernova~\citep{Heger_2002} in the atmosphere of a second-generation star indicating some Pop~III stars were more massive than $\sim 140$\,M$_{\odot}$~\citep{Xing_2023}. 
Furthermore, deep imaging from JWST of galaxies at $z > 10$ probes epochs early enough that the collective impacts of the first stars might be observable, with the $z = 10.6$ galaxy GN-z11 garnering particular attention due to the high equivalent width He\textsc{ii} emission lines from its halo~\citep{Maiolino_2023}, that potentially indicate Pop III stars with a top-heavy IMF. 
Intriguingly, it has also been argued that the presence of Pop~III stars with a top-heavy IMF may also explain the overabundance of UV-bright galaxies seen by JWST without resorting to alternatives to the standard $\Lambda$CDM cosmology~\citep{Naidu_2022, Harikane_2023, Trinca_2023}.

Measuring the Pop III IMF would thus provide insight into numerous outstanding questions in both astrophysics and cosmology.
21-cm cosmology promises to be a uniquely powerful probe of cosmic dawn, able to reveal the typical astrophysical population at high redshifts inaccessible to other observables~\citep{Furlanetto_2006}.
Consequently, it is anticipated to provide insight into the properties of the Pop III stars~\citep{Klessen_2023}, including, as we demonstrated in our previous paper~\citep{Gessey-Jones_2022}, the Pop III IMF.
In this paper, we show that \change{self-consistently including heating by Pop III XRBs significantly enhances the sensitivity of the 21-cm signal to the Pop III IMF}. 
As a result, we find that 21-cm global signal experiments already in operation, such as the Radio Experiment for the Analysis of Cosmic Hydrogen~\citep[REACH,][]{REACH}, and interferometers targeting the 21-cm power spectrum, such as the under-construction SKA-Low~\citep{SKA_Mission_Paper, Koopmans_2015}, are anticipated to be able to determine the Pop III IMF, even when accounting for the uncertainties in other high redshift astrophysics such as the properties of the metal-containing second generation of stars (Pop II) and the timing of the onset of Pop III star formation. \change{While previous studies have investigated the impact of Pop III high mass XRBs on the heating and ionization of the IGM~\citep[e.g.,][]{Xu2014}, and the observable 21-cm signature~\citep[e.g.,][]{Ahn2015, Mebane2020, Gessey-Jones_2022}, this work is the first to perform a self-consistent analysis to demonstrate the promising observational prospects for the Pop III IMF. }

\section{The Pop III IMF and the cosmological 21-cm signal}~\label{sec:21cm}

After recombination, but before reionization, neutral hydrogen atoms permeate the Universe~\citep[see][for review]{Furlanetto_2006}. 
These atoms have a hyperfine transition allowing them to emit or absorb 21\,cm wavelength photons. 
The net change in the radio background caused by this neutral hydrogen is the 21-cm signal, which is observed today at longer wavelengths thanks to cosmological redshift.
More concretely, we observe the signal from a patch of neutral hydrogen, at redshift $z$, as a differential brightness temperature~\citep{Madau_1997} at frequency $\nu = 1420/(1+z)$\,MHz with magnitude
\begin{equation}
T_{\rm 21}(z) = \left[1 - e^{- \tau_{\rm 21}(z)} \right] \frac{\left[ T_{\rm S}(z) - T_{\gamma}(z) \right]}{1 + z},
\label{eq1}
\end{equation}
where $\tau_{\rm 21}$ is the local optical depth of the 21-cm line, $T_{\gamma}$ is the radiation temperature of the radio background, and $T_{\rm S}$ is the statistical spin temperature which encapsulates the relative occupancy of the hyperfine states of neutral hydrogen. 
Since neutral hydrogen is abundant in the early Universe, the 21-cm signal is the most promising probe of the Universe at cosmic dawn.
As a result, various ongoing and under construction experiments~\citep[e.g.,][]{REACH, SKA_Mission_Paper, EDGES, SARAS3, HERA} are aiming to measure the 21-cm signal, principally via two summary statistics of the 21-cm signal field: the sky-averaged global 21-cm signal $\langle T_{\rm 21} \rangle(z)$, which encodes the temporal evolution of the 21-cm field (between $z \sim 30$ and $6$ for terrestrial experiments); and the 21-cm power spectrum $\Delta^2(k, z)$, which encodes the spatial variations in the field at wavenumber $k$ (typically targeting scales between $\sim 1$ and $1000$\,comoving Megaparsecs, cMpc).
The SKA-Low interferometer (see next section) may also be able to produce tomographic images of the $T_{\rm 21}$ field, though we do not consider this observable further in this study.

\begin{figure}[ht!]
    \centering
   \includegraphics{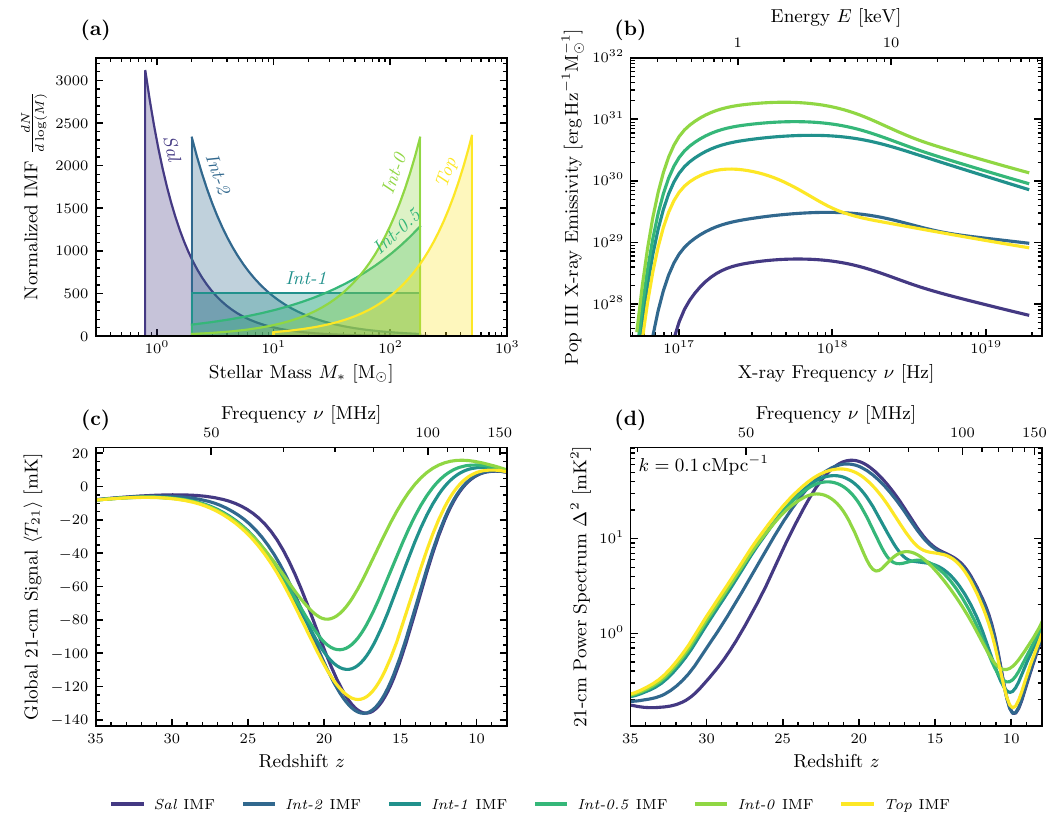}
   \caption{
   \textbf{Variation of the 21-cm signal with the Pop III IMF.} 
   \textbf{a}, The six truncated power-law IMFs we consider in this work. 
   \textbf{b}, Pop III X-ray emissivity computed for these IMFs. We observe significant variations between the emissivities, with a $257$ times difference in $f_{\rm X,III}$ (integrated luminosity in the range $0.20$ to $95.65$\,keV) and peak frequencies ranging from $0.92$ to $3.82$\,keV.
   \textbf{c},~We find the 21-cm global signal is sensitive to the Pop III IMF due to its impact on the X-ray emissivity and Lyman-band emissivity of Pop~III star-forming halos.
   \textbf{d}, A strong IMF dependence is also seen in the 21-cm power spectrum (shown at $k = 0.1$\,cMpc$^{-1}$). 
   IMF-induced variations in X-ray emissivity are the principal cause of the large difference in the 21-cm signals at $z < 23$, while changes in the Lyman-band emissivity of stars drive the impact at $z > 23$ (see \textit{Supplementary Materials} Fig.~\ref{fig:variation_origin}).
   To illustrate the impacts of the Pop~III IMF, here we only vary the IMF while keeping other astrophysical properties fixed \change{(see \textit{Methods} for details of the simulation code and its parameterization)}.
   } 
    \label{fig:variation_with_imf}
\end{figure}

The first stars play an essential role in shaping the 21-cm signal~\citep{Furlanetto_2006, Barkana_2016}.
At the end of the cosmic dark ages, \change{as the expansion of the Universe continues in the absence of stars, $T_{\rm S} \approx T_{\gamma}$ and so the 21-cm signal (Eq. \ref{eq1}) is anticipated to be small. However, as the first stars form, they emit Lyman-band radiation which enables the Wouthuysen-Field (WF) coupling~\citep{Wouthuysen_1952, Field_1958} in the surrounding IGM, causing $T_{\rm S}$ to rapidly approach the IGM temperature, $T_{\rm K}$. At these epochs, $T_{\rm K}$ is believed to have been lower than $T_{\gamma}$ as matter cools faster than photons in an expanding universe, and significant astrophysical heating has yet to occur. This leads to a rapid drop in the average 21-cm signal (Eq. \ref{eq1} and  Fig.~\ref{fig:variation_with_imf})) triggered by Pop III star formation  with the timing and strength of this feature sensitive to the Pop~III IMF~\citep{Gessey-Jones_2022}. This transition is also observed in the fluctuations of the 21-cm field, with the 21-cm power spectrum rising into a peak (see Fig.~\ref{fig:variation_with_imf}). }

After cosmic dawn, the WF coupling causes $T_{\rm S}$ to trace $T_{\rm K}$, thus making the 21-cm signal sensitive to any IGM heating mechanism. The first stars contribute to the heating of the IGM through both X-rays and their aforementioned Lyman-band radiation (Ly\,$\alpha$ heating).
Simulations show the former is dominant  \change{\citep{Reis_2021}}, with the principal source of these X-rays being X-ray binaries (XRBs), evolved binary systems wherein an astrophysical black hole or neutron star is accreting matter from a donor star that is overflowing its Roche lobe. Heating from Pop II and Pop III XRBs is anticipated to have been the principal driver raising the
temperature of the IGM after cosmic dawn \change{\citep{Fragos_2013, Fialkov_2014, Pacucci_2014, Gessey-Jones_2023}}, causing the 21-cm global signal to increase and potentially enter emission if
heating is vigorous enough. Similarly, the rising IGM temperature is reflected in the 21-cm power spectrum as an inflexion point or additional peak for scenarios with stronger heating (see Fig.~\ref{fig:variation_with_imf}).
The sensitivity of the cosmic dawn 21-cm signal to the Pop III IMF was previously explored via the subdominant Ly\,$\alpha$ heating~\citep{Gessey-Jones_2022} but not X-ray heating. 
A recent study~\citep{Sartorio_2023} has shown that the abundance and spectra of Pop III XRBs are highly sensitive to the IMF.
This sensitivity arises due to a complex interplay of factors, with the IMF impacting both the total number of Pop III binary systems and the mass of the stars within them.
The mass of the two Pop III stars in the binary, in turn, determines whether the primary star becomes a black hole or neutron star and the duration for which, if at all, the secondary star undergoes Roche-lobe overflow.
Hence, we anticipate X-ray heating will enhance the signature of the Pop~III IMF in the 21-cm signal.

To self-consistently model the sensitivity of the 21-cm signal to the Pop III IMF thus requires including all of the above mechanisms.
The impact of Lyman photon-mediated effects ~\citep[see \textit{Methods} and][]{Gessey-Jones_2022} are already modelled in the semi-numerical code \textsc{21cmSPACE}~\citep[\change{21-cm Semi-numerical Predictions
Across Cosmic Epochs}, e.g.][]{Visbal_2012, Fialkov_2014, Gessey-Jones_2023}.
Hence, we extend this code by incorporating the IMF-dependent contribution of Pop III XRBs to heating and ionization. 
Our X-ray spectra are calculated using state-of-the-art population synthesis simulations, which sample stellar binaries from the chosen IMF, evolve the metal-free binary systems using \textsc{binary\_c} \change{\citep[e.g.][]{Izzard_2023,Mirouh_2023}} to identify systems that become XRBs, determine the evolving spectra of the XRBs using a multi-colour thin disc model\change{, and apply a halo mass-dependent X-ray escape fraction. See \textit{Methods}~\ref{ssec:xrb}~\citep[and][]{Sartorio_2023} for more details on the assumptions and the uncertainties of these simulations, including the prescriptions utilized for supernova kicks, the common-envelope phase, and mass transfer.} 

Due to the computational cost of simulating XRB catalogues, we are limited to considering a selection of six IMFs (see Fig.~\ref{fig:variation_with_imf}\change{, a}) in this study.
We consider truncated power-law IMFs of the form
\begin{equation}~\label{eqn:imfs}
    \frac{dN}{dM} \propto M^{-\alpha_{\rm III}}, \qquad M \in [M_{\rm min}, M_{\rm max}],
\end{equation}
with the parameters listed in Table~\ref{tab:imfs}.
These IMFs are chosen to encapsulate our large uncertainties~\citep{Stacy_2013, Hirano_2014, Hirano_2015, Wollenberg_2020, Prole_2022, Jaura_2022,Klessen_2023}, including extreme IMFs dominated by  low-mass  (\textit{Sal}, $\alpha_{\rm III} = 2.35$) and high-mass stars (\textit{Top}, $\alpha_{\rm III} = 0$), as well as investigating the impacts of the transition from a bottom-heavy to a top-heavy IMF (while keeping the minimum and maximum stellar mass fixed).

We show the specific X-ray emissivity calculated for each Pop III IMF in Fig.~\ref{fig:variation_with_imf} \change{(b)} and list the corresponding values for the integrated emissivity between $0.20$ and $95.65$\,keV, $f_{\rm X, III}$, in Table~\ref{tab:imfs}.
We find a strong dependence of specific X-ray emissivity on the IMF, with $f_{\rm X, III}$ varying by up to a factor of $257$ between the IMFs.
Bottom-heavy ($\alpha_{\rm III} \geq 2$) IMFs are found to have the lowest emissivities.
This is due to a combination of having a relatively small fraction of massive stars that can form the black holes or neutron stars necessary for XRBs~\citep{Sartorio_2023}, and the resulting XRBs having less massive accretors and thus lower luminosities. The extremely top-heavy \textit{Top} IMF is also found to have a moderate integrated luminosity. This is due to a smaller number of stars (and, hence, binaries), and the XRBs produced being shorter-lived. 
The largest integrated X-ray luminosities instead occur for intermediate IMFs that strike a balance between these effects~\citep[discussed further in \textit{Methods} and][]{Sartorio_2023}.

\bgroup
\def\arraystretch{1.5}
\begin{table}[ht!]
    \centering
    \begin{tabular}{|c|c|c|c|c|c|c|c|}
    \cline{1-2}\cline{4-6}\cline{8-8}
         \multirow{2}{*}{IMF Name }  & Color &&  IMF exponent & Minimum stellar& Maximum stellar  & & Relative specific X-ray \\[-5pt]
         & code & & $\alpha_{\rm III}$  & mass $M_{\rm min}$ (M$_{\odot}$)  & mass $M_{\rm max}$ (M$_{\odot}$) & & emissivity $f_{\rm X, III}$\\
         \cline{1-2}\cline{4-6}\cline{8-8}
         \textit{Sal}  (i.e., Salpeter~\citep{Salpeter_1955}) & \tikz\draw[black,fill=VirdisA, anchor=center] (0,0) rectangle ++(3em,1.4ex); &  & 2.35 & 0.8 & 250.0 & & 0.29 \\
         \cline{1-2}\cline{4-6}\cline{8-8}
         \textit{Int-2}   & \tikz\draw[black,fill=VirdisB, anchor=center] (0,0) rectangle ++(3em,1.4ex);& & 2.00 & 2.0 & 180.0 & & 2.85\\
         \cline{1-2}\cline{4-6}\cline{8-8}
         \textit{Int-1}& \tikz\draw[black,fill=VirdisC, anchor=center] (0,0) rectangle ++(3em,1.4ex); && 1.00 & 2.0 & 180.0 & & 35.27\\
         \cline{1-2}\cline{4-6}\cline{8-8}
         \textit{Int-0.5} &\tikz\draw[black,fill=VirdisD, anchor=center] (0,0) rectangle ++(3em,1.4ex); & & 0.50 & 2.0 & 180.0 & & 46.40\\
         \cline{1-2}\cline{4-6}\cline{8-8}
         \textit{Int-0} & \tikz\draw[black,fill=VirdisE, anchor=center] (0,0) rectangle ++(3em,1.4ex); &  & 0.00 & 2.0 & 180.0 & & 75.24\\
         \cline{1-2}\cline{4-6}\cline{8-8}
         \textit{Top} & \tikz\draw[black,fill=VirdisF, anchor=center] (0,0) rectangle ++(3em,1.4ex); &  & 0.00 & 10.0 & 500.0 & & 3.21\\
         \cline{1-2}\cline{4-6}\cline{8-8}
    \end{tabular}
    \caption{\textbf{Pop III stellar IMFs.} 
    Each IMF is a truncated power-law with minimum mass $M_{\rm min}$, maximum mass $M_{\rm max}$, and exponent $\alpha_{\rm III}$ (see \cref{eqn:imfs}).
    The \textit{Sal}~\citep{Salpeter_1955} and \textit{Top} IMFs represent intentionally extreme bottom-heavy and top-heavy scenarios, respectively, and the four \textit{Int} IMFs allow us to explore the effects of varying the IMF slope while keeping the boundaries fixed.
    \change{In this study we consider Pop III stars between $0.8$ and $500$\,M$_{\odot}$.
    The lower limit is motivated by stellar archaeology surveys~\citep{Magg_2019} finding Pop III stars below this mass threshold must have been very rare. The upper limit is set to  $500$\,M$_{\odot}$ because more massive stars experience early photodisintegration of their atmospheres, resulting in us being unable to model their spectra reliably~\citep{Gessey-Jones_2022}.}
    Also given is $f_{\rm X, III}$, the computed X-ray emissivity of Pop III star-forming halos per unit star formation rate for each IMF (see \textit{Methods}); for convenience, these values are normalized so that $f_{\rm X, III} = 1$ corresponds to  $3\times 10^{40}$\,erg\,s$^{-1}$\,M$_{\odot}^{-1}$\,yr, the specific X-ray emissivity expected for Pop II star-forming halos based on low-redshift observations and simulations~\citep{Fragos_2013b}. 
    }
    \label{tab:imfs}
\end{table}
\egroup

In Fig.~\ref{fig:variation_with_imf} \change{(c and d)}, we depict the variation in the 21-cm global signal and power spectrum caused by changing the Pop III IMF.
To illustrate the impacts of the IMF, here we keep the other parameters of \textsc{21cmSPACE}, including cosmology, Pop II star \change{and XRB} properties, the Pop III star formation efficiency\change{,} and the duration of the Pop II to Pop III transition, fixed to fiducial values (see \change{\textit{Supplementary Materials} Table~\ref{tab:priors}}).
As in our previous work~\citep{Gessey-Jones_2022}, differences in the 21-cm signals are seen as early as $z\sim30$ due to the Lyman-band effects. 
The bottom-heavy IMFs have a later onset of the 21-cm global signal absorption trough, as well as a later and stronger peak in the 21-cm power spectrum at cosmic dawn. This is because low-mass stars dominating bottom-heavy IMFs have less efficient Lyman-band emission, resulting in a delayed WF effect, weaker Ly\,$\alpha$ heating, and reduced Lyman-Werner feedback. 

Owing to the new addition of the IMF-dependent Pop III XRB spectra, which affect the IGM heating, we see considerable variation in the 21-cm signals at $z = 10-23$, not seen in our previous study~\citep{Gessey-Jones_2022}.
We find variations of up to $56$\,mK and $\Delta z \sim 3$ in the depth and timing of the absorption trough, differences in the power spectrum at $k=0.1$\,cMpc$^{-1}$  and $z = 20$ of $> 70$\,mK$^2$, and an additional peak appearing in the 21-cm power spectra around $z = 15$ for some IMFs (e.g., \textit{Int-0}).
These novel effects are caused by the IMF-dependent intensity of X-ray emissivity (e.g., the factor of $257$ spread in $f_{\rm X, III}$ values), with the variations in emissivity shape having a subdominant impact (see \textit{Supplementary Information}).  
This also explains why the three IMFs with lower emissivities (\textit{Sal}, \textit{Int-2}, and \textit{Top}) have similar 21-cm signals in this redshift range, as in these cases, Pop III X-ray emission is inefficient and hence the heating of the IGM is being principally driven by Pop~II stars instead.
Thus, due to X-ray heating, the Pop III IMF has a significantly stronger impact on the 21-cm signal over a wider redshift range than previously anticipated.

The details and the strength of the variations in the 21-cm signal as a function of \change{the Pop III} IMF will depend on other astrophysical processes, which remain uncertain\change{.
For example, cosmic ray heating \citep[][]{Gessey-Jones_2023} from Pop II and Pop III supernovae may also be efficient, in which case the signature of the Pop III IMF would be expected to be further enhanced due to intrinsic links between the rate and types of supernovae and the IMF. Despite these uncertainties, } the above examples are sufficient to demonstrate that the Pop III IMF can significantly impact the 21-cm signal. This realization has two major consequences for 21-cm cosmology. 
First, measurements of the 21-cm signal can potentially be used to constrain the IMF of the first generation of stars. 
Secondly, assuming an incorrect IMF model could bias the inference of other astrophysical and cosmological information from the 21-cm signal (e.g., the nature of dark matter~\citep{Barkana_2018}). 
A realistic assessment of the significance of these constraints and estimates of biases requires the specification of expected experimental sensitivities, which we will address next.

\section{Experiments, Synthetic 21-cm Signal Measurements, and Data Analysis} \label{sec:experiments}

The field of observational 21-cm cosmology is rapidly evolving with ongoing experiments ~\citep[e.g.][]{MIST, SARAS3, HERA, LOFAR, NenuFar, MWA} progressively improving upper limits, and the under-construction SKA  scheduled for commissioning in 2028. Here we consider prospective constraints on the Pop III IMF using the sensitivities of the operational REACH global signal experiment and SKA as examples.    

REACH \citep{REACH} is a state-of-the-art experiment to detect the 21-cm global signal using multiple radiometers and a comprehensive Bayesian analysis pipeline~\citep{Anstey_2021}.
At the time of writing, REACH is undertaking its first phase of observations. 
In their mission paper~\citep{REACH}, the REACH collaboration outlines three potential scenarios for their final global 21-cm signal measurement (post foreground removal) with sensitivities of $250$ (pessimistic), $25$ (anticipated) or $5$\,mK (optimistic) at a $0.1$\,MHz resolution across the redshift band of $z = 7.5-28.0$.
They additionally assume these measurements are noise-limited, modelling this as white Gaussian noise; hence, the above is equivalent to $79.1$/$7.9$/$1.6$\,mK sensitivity when integrated into $1$\,MHz bins.
In our subsequent analysis, we shall consider the same three cases and adopt the same assumptions.
Due to these inherited assumptions, our REACH forecasts represent a best-case scenario where foreground removal does not produce a significant systematic uncertainty in the measured signal~\citep{Heimersheim_2024}; to help achieve this REACH proposes to use multiple antennas and a time-separated analysis which exploits the local sidereal time invariance of the global 21-cm signal~\citep{Anstey_2023}.

SKA-Low is the low-frequency radio interferometer part of the SKA observatory~\citep{SKA_Mission_Paper}.
A principal goal of the interferometer is to detect the 21-cm power spectrum.
To model future measurements by this experiment, we use existing predictions of the sensitivity of $300$, $1000$, and $3000$ hours of observations by the SKA-Low core~\citep{Koopmans_2015} in the $z = 7.0-27.0$ redshift band, and achromatic Gaussian noise. 
We simulate a foreground avoidance strategy by considering measurements only at $k > 0.1$\,cMpc$^{-1}$ and are limited to $k \leq 1.0$\,cMpc$^{-1}$ by experimental resolution.
Due to us only considering the spherical 21-cm power spectrum measurement by the SKA-Low core in a limited wavenumber range, our forecasts for this experiment will be quite conservative. 
Consequently, our subsequent SKA-Low constraints should be treated as lower bounds on those that the full experiment will achieve within a given observation time, being improved upon in practice via the additional information made available by the cylindrical 21-cm power spectrum~\citep{Prelogovic_2024}, foreground mitigation, and access to higher wavenumbers.

At this juncture, we should discuss the disputed detection of the 21-cm global signal by the EDGES experiment~\citep{EDGES, Hills_2018, Sims_2020}.
If it is of cosmological origin, its anomalous depth would suggest exotic physics that enhances the 21-cm signal, such as dark matter cooling~\citep[e.g.][]{Barkana_2018} or a strong excess radio background~\citep[e.g.][]{Feng_2018}.
Any physics that enhances the 21-cm signal would also increase the 21-cm signal differences between Pop III IMFs, reducing the sensitivity required for constraints. 
However, the EDGES detection may not be cosmological~\citep{Hills_2018, Sims_2020}. 
As a result, we do not attempt to use the EDGES data to constrain the Pop~III IMF in this study. 
Furthermore, to ensure a conservative assessment of the Pop~III IMF constraints, we include only a weak astrophysical radio background that is negligible compared to the CMB~\citep[$< 2.8$\% at all redshifts,][]{Sikder_2023} and do not include dark matter cooling when simulating the 21-cm signal. We also do not consider the \change{existing} bounds on the 21-cm global signal from SARAS~3~\citep{SARAS3} or the power spectrum limits from HERA~\citep{HERA_theory_22} as they only slightly disfavour high Pop III star formation efficiencies~\citep{Pochinda_2023}, and so are not yet sufficiently constraining to provide insight into the IMF.

To make our forecasts as realistic as possible, we perform them by generating and analysing synthetic experimental measurements. 
The synthetic measurements combine 21-cm signals generated using \textsc{21cmSPACE} and noise added of the appropriate level and form for each  experiment and sensitivity.
We adopt a nested-sampling-based Bayesian methodology for the analysis of this data, inspired by the approach taken by REACH~\citep{REACH}.
This methodology allows us to rigorously marginalize over uncertain high-redshift astrophysical parameters such as the Pop III star formation efficiency and properties of Pop II stars \change{and XRBs} (see \change{\textit{\textit{Supplementary Materials}} Table~\ref{tab:priors}} for a full list of the parameters we fit for and their intentionally broad priors).
Such an approach is essential to guarantee our forecasts properly account for any degeneracies between the impacts of the Pop III IMF and those of other astrophysics on the 21-cm signal \change{(discussed further in Methods~\ref{ssec:simulations})}.

As a single nested sampling run would require millions of 21-cm signal simulations, \textsc{21cmSPACE} is too slow for direct use in this analysis due to its approximately three-hour runtime. 
To remedy this issue, we create neural network emulators of \textsc{21cmSPACE} and use these in our likelihoods and synthetic data generation, following the methodology applied to the analysis of the existing HERA and SARAS 3 data \change{in our earlier work} ~\citep[e.g.][]{Bevins_2022}.
For more details on the synthetic data generation and analysis, see \textit{Methods}.

\section{Results} \label{sec:results}

Although the cosmic Pop III IMF is highly uncertain, to generate a synthetic measurement we need to assume a \textit{data IMF}, which serves as the ground truth when analyzing that data set.
Unless otherwise stated, we adopt the \textit{Int-1} IMF as the data IMF as it is consistent with current theoretical expectations~\citep{Klessen_2023}.

\begin{figure}[ht!]
    \centering
   \includegraphics{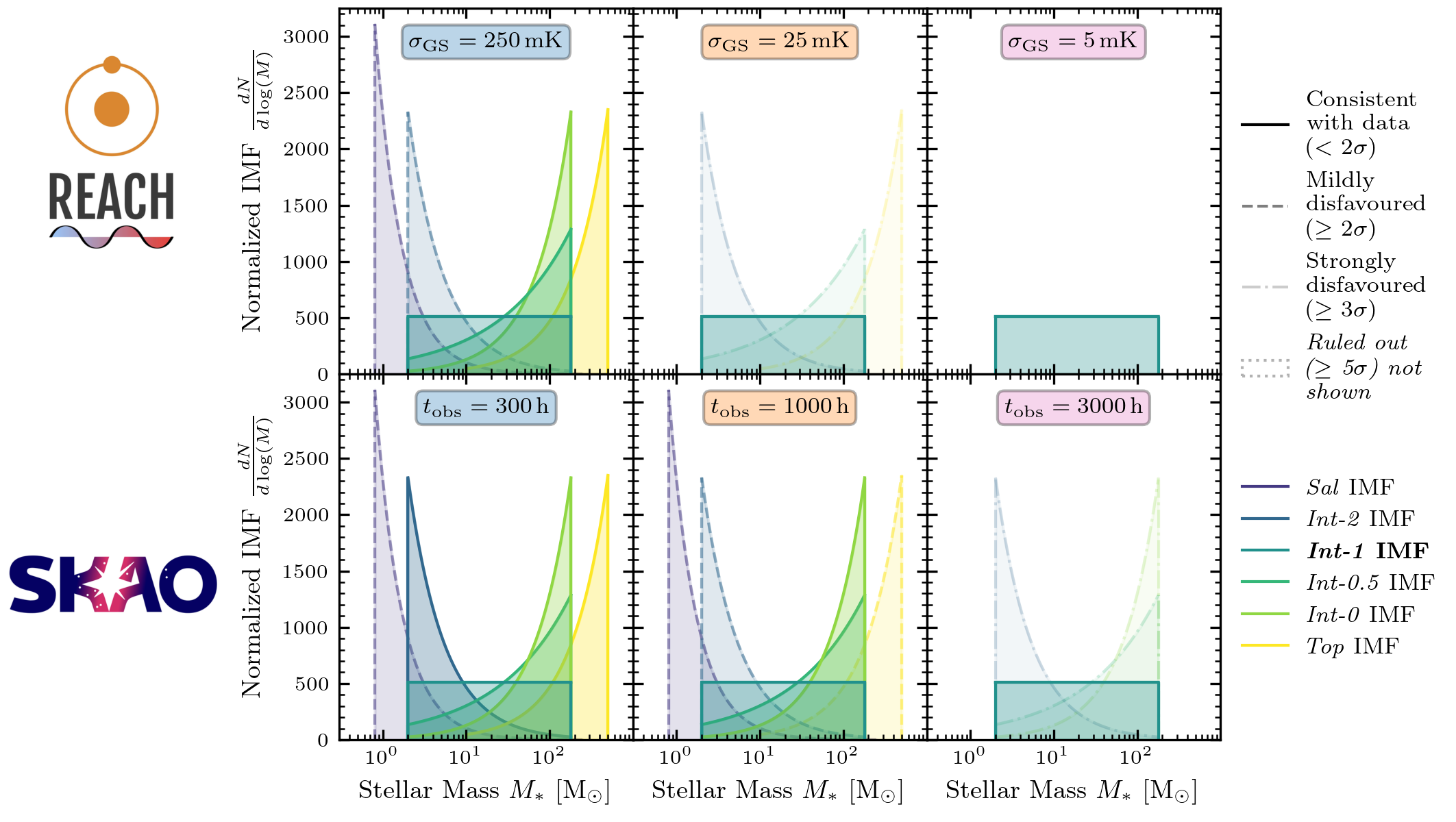}
   \caption{
   \textbf{Prospective constraints on the mass distribution of the first stars from 21-cm signal experiments.} 
For this figure, the synthetic measurement data was generated using the \textit{Int-1} IMF indicated in bold in the legend.
Each panel shows the six Pop III IMFs we consider in this study (see Fig.~\ref{fig:variation_with_imf} and Table~\ref{tab:imfs}), with the marginalized \textit{a posteriori} confidence in each IMF indicated via its line type and opacity. 
The top row shows prospective IMF constraints from the 21-cm global signal experiment REACH at $250$, $25$, and $5$\,mK sensitivity (from left to right), and the bottom row shows constraints from the 21-cm power spectrum experiment SKA-Low with $300$, $1000$, and $3000$\,h of observations (also from left to right).
REACH at $25$\,mK and SKA-Low at $3000$\,h are both found to be able to identify the correct IMF (\textit{Int-1}) at $> 3$\,$\sigma$ confidence, showing the potential of the 21-cm signal to determine the Pop III IMF.
   } 
    \label{fig:headline_results_plot}
\end{figure}

We depict the prospective constraints (posterior distributions) on the Pop~III IMF in Fig.~\ref{fig:headline_results_plot} for REACH and SKA-Low at different sensitivities.
In all cases, we find \textit{Int-1} is always correctly identified as being consistent with the data, and the five alternative IMFs (e.g., non-data IMFs) are disfavoured at increasing statistical significance as sensitivity improves. 
For REACH at $250$\,mK sensitivity, we find only mild disfavouring ($\geq 2$\,$\sigma$) of the two IMFs that have the weakest Pop III contribution to X-ray heating (\textit{Sal} and \textit{Int-2} IMF) and hence on average produce the strongest, and thus most easily ruled-out, 21-cm signals. 
At $25$\,mK all the alternative IMFs are disfavoured at $> 3$\,$\sigma$, with the significance increasing to $> 5$\,$\sigma$ at $5$\,mK. SKA-Low would only mildly disfavour one alternative IMF (\textit{Sal}) with 300\,h of observations. The situation improves as we gradually increase the observational time. We find that with $1000$\,h, we can disfavour three alternative IMFs, while $3000$\,h of SKA-Low observations allow us to detect the correct data IMF at  $>3$\,$\sigma$, with some alternative IMFs ruled out at higher significance ($> 5$\,$\sigma$). 
In \textit{Supplementary Materials}, we show the equivalent results for when each one of our six IMFs is taken to be the data IMF, finding that the data IMF is always consistent with the synthetic measurement, and REACH at $25$\,mK sensitivity, or SKA-Low at $3000$\,h of integration is sufficient to constrain the IMF at $> 3$\,$\sigma$.

In addition to the forecasts for individual experiments, we considered a joint analysis between REACH at $25$\,mK sensitivity and $1000$\,h of SKA-Low observations (see \textit{Supplementary Materials} for other combinations).
As expected, the joint analysis improved the IMF constraints over the individual experiments, with a factor of $> 6$ decrease in the posterior probability of all alternative IMFs. These findings motivate joint analyses using future data to increase the statistical significance of any 21-cm cosmology constraints on the Pop III IMF.

Our methodology also allows us to demonstrate the biases introduced by assuming an incorrect Pop III IMF when analysing 21-cm signal data (see Fig.~\ref{fig:biases}).
We find that the impacts of the Pop III IMF on the 21-cm signal can be partially compensated for by varying other cosmic dawn and reionization astrophysical parameters, resulting in the values inferred for these other parameters being biased.
These biases are found to increase with experimental sensitivity; for example, in a $25$\,mK global 21-cm measurement, we see $> 2$\,$\sigma$ biases in some Pop II star properties which rises to $> 5$\,$\sigma$ biases in a $5$\,mK measurement. 
A consideration of the uncertainties in the Pop III IMF will thus be necessary to ensure accurate interpretation of future precision 21-cm signal measurements.

\section{Discussion and Conclusions}\label{sec:conclusions}

We have explored the sensitivity of the 21-cm signal to the Pop~III IMF, finding that previous analyses have significantly underestimated it\change{~\citep{Gessey-Jones_2022}}. 
This is because \change{those} analyses neglected the dependence of X-ray heating on the Pop III IMF via the abundance and luminosities of Pop III XRBs.
Due to this greater sensitivity, we predict that 21-cm signal experiments should reveal the mass distribution of these elusive first stars.
We find that $> 3$\,$\sigma$ significance constraints on the Pop III IMF can be achieved by either a noise-limited measurement of the 21-cm global signal at 25\,mK sensitivity \change{(e.g., with REACH)}, or by 3000\,h of SKA-Low 21-cm power spectrum observations.
A joint analysis of these two observables can further improve these constraints. 

Our conclusions are \change{found to be} robust to the uncertainty in Pop III IMF and were derived using a nested-sampling Bayesian methodology to \change{account for the potential degeneracies with other poorly-constrained high-redshift astrophysics such as Pop III and Pop II star formation efficiencies. 
Although we do not include all possible stellar mass-dependent effects, e.g. cosmic ray heating and gradual metal enrichment are omitted for simplicity, their self-consistent modelling is expected to further enhance the dependence of the 21-cm signal on the Pop III IMF, thus strengthening our conclusions. 
Furthermore, while there remain uncertainties in Pop III XRB formation, we anticipate these to be either IMF-independent or have a smaller impact on the IGM X-ray heating and the 21-cm signal than the IMF and thus have a low impact on our forecast constraints.}

A noise-limited measurement of the global 21-cm signal at $25$\,mK may prove difficult due to the challenge of removing the significant radio foregrounds.
However, our SKA-Low forecasts were intentionally conservative.
They employed a foreground avoidance strategy, considered only the array core, and used the spherical 21-cm power spectrum rather than the cylindrical power spectrum. 
Consequently, the full SKA-Low will be able to achieve stronger IMF constraints in less observation time than our estimates above.
With SKA-Low commissioning expected in 2028 and other experiments in progress attempting to make conclusive global signal~\citep[e.g.][]{MIST, SARAS3} or power spectrum detections~\citep[e.g.][]{HERA, LOFAR, NenuFar, MWA}, the prospect for 21-cm cosmology to measure the masses of the first stars in the coming years is very promising. 

The insights gained from a measurement of the Pop III IMF via the 21-cm signal will shed light on cosmic dawn, a critical and largely uncertain transition in cosmological history.
Furthermore, this measurement will elucidate high-redshift galaxy observations by independently testing the proposals that Pop III stars with a top-heavy IMF explain certain JWST observations~\citep{Harikane_2023, Trinca_2023, Maiolino_2023} and aid in the interpretation of the $z > 10$ black hole-black hole merger events targeted by next-generation gravitational wave experiments~\citep{CosmicExplorer}.
Thus, our findings provide an additional compelling case for 21-cm cosmology.

\setcounter{section}{0}
\renewcommand\thefigure{M.\arabic{figure}}
\renewcommand\thetable{M.\arabic{table}}
\renewcommand\thesubsection{M.\arabic{subsection}}
\renewcommand\thesection{}

\newpage
\section*{Methods} \label{sec:method}

\subsection{Modelling of Pop III stellar spectra}\label{ssec:spectra}

In our 21-cm signal simulations, we utilise the Pop III star spectra computed for our previous paper~\citep{Gessey-Jones_2022}.
Here, we summarise the methodology of that work for completeness (see the original paper for further details and a discussion of the spectra). 

The Pop III star spectra were calculated by combining stellar evolution tracks with a grid of precomputed stellar spectra.
These stellar evolution tracks were simulated~\citep{Mirouh_2023} using the Modules for Experiments in Stellar Astrophysics (\textsc{mesa}) stellar evolution code~\citep[][version 12115]{MESA_I, MESA_II, MESA_III, MESA_IV, MESA_V}, assuming the stars evolved in isolation (discussed further at the end of this section), were non-rotating, were initially metal-free ($Z = 0$), and had negligible mass-loss. 
The tracks begin at the zero-age main sequence and end at core hydrogen depletion ($M < 310$\,M$_{\odot}$) or photo-evaporation ($M > 310$\,M$_{\odot}$).  
Meanwhile, the stellar atmosphere modelling code \textsc{tlusty}~\citep[][version 205]{TLUSTY0, TLUSTYI, TLUSTYII, TLUSTYIII} was utilised to calculate stellar spectra on a grid of effective temperature and surface gravity values.
This grid was designed to cover the aforementioned stellar evolution tracks.
Atmospheres were calculated using non-local thermodynamic equilibrium modelling with atomic and ionic lines included~\citep{Auer_1969}.
\change{Chemically, the atmospheres were modelled as being composed of only hydrogen and helium in Big Bang nucleosynthesis proportions.
This is motivated by the expectation that Pop~III stars formed metal-free and our stellar evolution simulations predicting negligible amounts of the metal synthesised within the stars reaches their surface during their main sequence.}

The lifetime spectra of individual Pop III stars were then calculated by integrating along each stellar evolution track, interpolating the grid of stellar spectra to each point on the evolution track. 
These individual stellar spectra were subsequently incorporated into \textsc{21cmSPACE} (see section \ref{ssec:simulations}), with the Pop III population-averaged Lyman-band (and Lyman-Werner band) emissivity finally computed at runtime by integrating these spectra weighted appropriately by the Pop III IMF.

A recent study~\citep{Tsai_2023} found that binary evolution and mass transfer increase the total UV emission of Pop III binaries when compared to the same stars evolving in isolation.
By combining, their worst-case results (largest difference) with predictions for the Pop III interacting binary fraction~\citep{Tsai_2023, Stacy_2013}, we can estimate that not including these mechanisms has led to the Pop III Lyman-Werner emissivities used in \textsc{21cmSPACE} being underestimated by, at most, $25\%$.
Hence, while not modelling a portion of Pop III stars as binaries when calculating Lyman-band emissivities is somewhat inconsistent with the XRB modelling of the next section, the impact of this assumption is small.

\subsection{Modelling of Pop III XRBs}\label{ssec:xrb}

Our modelling of Pop III XRBs follows a two-step methodology (originally described in~\cite{Sartorio_2023}).
In the first step, we create catalogues of metal-free ($Z = 0$) binaries at high redshifts for all six explored IMFs. 
To do this, we take \textsc{21cmSPACE} simulations of the Pop III star formation rate and sample them according to the IMF to create a catalogue of stars forming at each redshift.
We then compute the number of newly formed binary systems by assuming 28\% of Pop III stars are in binaries \change{(note, other works sometimes instead quote the fraction of stellar systems that are binaries)}.
The utilized value for binary fraction is intentionally conservative, being approximately half the value found in \change{some of the} hydrodynamic simulations of Pop III protostar multiplicity~\citep[][]{Stacy_2013, Riaz_2018, Sugimura2020}. \change{As we do not use a mass-dependent pairing of stars in a binary system, any error in the assumed binary fraction should scale the X-ray emissivities of each IMF by the same factor. 
Thus, the current uncertainty in Pop III binary fractions introduce at most a factor of three variation in the predicted Pop III X-ray emissivity.
Consequently, this uncertainty does not strongly impact our forecast constraints as we find the variation in the total Pop III X-ray emissivity between IMFs to be $\mathcal{O}(300)$ (Table~\ref{tab:imfs}).}
With the number of binaries established, we assign a subset of the sampled stars to those systems, pairing them randomly. 
Then, we set the initial orbital parameters for each binary system by Monte-Carlo sampling of mass-dependant observationally motivated probability distribution functions~\citep{Jeans_1919, Kroupa_1995, Sana_2012}.

In the second step, the catalogue of binaries is fed as an initial condition to the population synthesis code \textsc{binary\_c}~\citep{Izzard_2004, Izzard_2006, Izzard_2009, Izzard_2018, Izzard_2023, Mirouh_2023}.
For self-consistency in our \textsc{binary\_c} simulations, we employ the same metal-free stellar evolution tracks as we used to compute the Pop III stellar spectra (see previous section).
A binary is classified as an XRB whenever the primary is a black hole or neutron star and is accreting material from a companion via winds or Roche-lobe overflow. 
We then calculate the XRB's spectral energy distribution (SED) assuming a thin accretion disk model, taking into account Comptonization by a corona, the accretion rate, the mass of the primary compact object, and the binary's orbital parameters. 
\change{We model the X-ray escape fraction from the host halo \citep[following][]{Sartorio_2023} by assuming halo-mass dependent absorption and adopting primordial gas composition (76\% of atomic hydrogen and 24\% of helium).} The X-ray emission of each XRB is assembled by \textsc{binary\_c} into a catalogue of XRB lifetimes and energy emission across the frequency range $4 \times 10^{-8}$ to $4 \times 10^{10}$\,keV. 
Additionally, we compute the local absorption of X-rays by hydrogen and helium within dark matter halos in which the XRBs are situated to obtain the X-ray spectrum that would escape and affect the IGM.

Finally, we normalize the total escaping X-ray emissivity using the simulated star formation rate density to find the specific X-ray emissivity from Pop III star-forming haloes. 
These population-averaged emission spectra (shown in Fig.~\ref{fig:variation_with_imf}) were then integrated into \textsc{21cmSPACE} as described in the next section.

For 21-cm cosmology, the relevant part of the X-ray spectrum is from $0.1$ to $2.5$\,keV.
X-rays at these energies can readily escape into the IGM~\citep{Sartorio_2023} but are still absorbed in a Hubble time~\citep{Furlanetto_2006} and thus contribute to heating the IGM.
We find a large portion of the X-ray emission of Pop III XRBs lies in this range, with the peak of the X-ray emissivities either lying within this band 
($0.9$, $2.0$, $2.4$, and $2.4$\,keV for the \textit{Top}, \textit{Int-0}, \textit{Int-0.5}, and \textit{Sal} IMFs respectively)
or just above it
($3.0$\,keV for the \textit{Int-1} IMF, and $3.8$\,keV for the \textit{Int-2} IMF). 
Hence, based on the spectra alone, we anticipate Pop III XRBs could efficiently contribute to the X-ray heating of the IGM \change{\citep[in agreement with literature, e.g., ][]{Xu2014} with the signature of this heating being IMF-dependent}.

Alongside these variations in peak frequency, we observe the integrated specific X-ray emissivity $f_{\rm X, III}$ (Table~\ref{tab:imfs}) differs by a factor of $257$ between IMFs (as highlighted in the main text).
The sensitivity of X-ray emissivity to the Pop III IMF stems from the interplay of four key mechanisms~\citep{Sartorio_2023}: the number of Pop III stellar binaries that form, the proportion of binaries that become XRBs, the lifetime of the XRBs, and the spectra of the individual XRBs.
Bottom-heavy IMFs lead to more Pop III stars and stellar binaries; however, these stars are generally of lower mass, and thus few of these binaries have primaries large enough to become the neutron stars or black holes required to form an XRB. 
Hence, the proportion of binaries that become XRBs tends to increase as the IMF becomes more top-heavy (this is somewhat complicated by pair-instability supernovae causing some massive Pop III stars to leave no remnant behind ~\citep{Heger_2002}). 
In addition, the typical primary star mass increases as the IMF becomes more top-heavy, leading to a greater accretion rate and more efficient X-ray emission, producing individual XRBs that are more luminous on average. 
However, the typical XRB lifetime simultaneously decreases due to the increasing mass of the secondary star.  
Thus, we see these four competing effects, with the number of binaries forming and XRB lifetime being highest for bottom-heavy IMFs and the proportion of binaries that become XRBs and XRB luminosity being highest for top-heavy IMFs. 
The balance of these mechanisms leads to the trends seen in Fig.~\ref{fig:variation_with_imf} and Table~\ref{tab:imfs}, where the intermediate IMFs have the highest X-ray emissivities, with extremely top-heavy or bottom-heavy IMFs having lower emissivities.
A more detailed discussion of these effects and their impacts on the X-ray emission properties of Pop III star-forming halos can be found in~\citep{Sartorio_2023}.

\change{Some aspects of Pop III XRB formation and evolution, such as supernova kicks, common-envelope evolution, and mass accretion, are poorly understood and thus could present an obstacle to definitively determining the Pop III IMF from the 21-cm signal. Robustly assessing the impact of each of these uncertain processes demands quantifying their effects on the 21-cm signal and propagating this through a forecasting analysis, which is beyond the capabilities of state-of-the-art simulations and the scope of this paper. However, we argue that the uncertainties mentioned above are expected to have a smaller impact on the observable signal than the Pop III IMF. Thus, while they may somewhat weaken our constraints on the Pop III IMF, they are unlikely to eliminate them. First, we have used \textsc{binary\_c} to test several supernova kick distributions based on observational data of pulsar velocities \citep{Arzoumanian2002, Hobbs2005} finding the impact on the number of XRBs is small, being at most 10\%. Second, the effect of differences in common-envelope evolution on Pop II XRBs has previously been explored in the literature with a strong effect reported on low-mass XRBs; in contrast, high-mass XRBs were found to be largely unaffected \citep{Fragos_2013b}. Although an analogous study of Pop III XRBs does not exist, we expect a low impact on our results as the majority of Pop III XRB X-ray emission in our work is due to high-mass XRBs. Finally, as Pop III XRBs are overwhelmingly powered by Roche-lobe overflow (the absence of metals makes line-driven winds very inefficient), changes in mass transfer efficiency scales the average luminosity of Pop III XRBs in an IMF-invariant manner. Hence, like the binary fraction, this introduces an IMF-independent scaling factor in the X-ray luminosity that is likely much smaller than the $\mathcal{O}(300)$ variation between IMFs.
}

\subsection{Simulations of the 21-cm signal}\label{ssec:simulations}

To simulate the 21-cm signal, we utilize \textsc{21cmSPACE}  \citep[][]{Visbal_2012, Fialkov_2012, Fialkov_2013, Fialkov_2014, Cohen_2016, Fialkov_2019, Reis_2020, Reis_2021, Reis_2022, Magg_2022, Gessey-Jones_2022, Gessey-Jones_2023, Sikder_2023}. 
The core idea behind the code is a separation of scales between large-scale phenomena, like radiative transfer and the variations in the overdensity of the Universe, and small-scale phenomena, like dark-matter halo collapse and star formation. 
These small-scale phenomena are not resolved in the simulations; instead, they are modelled using analytic approximations or fitting formulas of finer-resolution simulations. 
Conversely, large-scale phenomena are resolved, with the simulation volume split into a grid of $128^3$ cells of side-length $3$\,cMpc. 
Such large cells are necessary for three main reasons: to ensure cells remain in the linear growth regime throughout the simulation, to give sufficient cell volume for some of the aforementioned analytic models to be valid, and to allow for simulation volumes large enough for accurate statistics on the cosmological observables of interest while keeping the simulation runtime reasonably short.

\textsc{21cmSPACE} models a wide range of physics known to impact the 21-cm signal including, but not limited to: Pop~III and Pop~II star formation and radiative emission~\citep{Magg_2022, Gessey-Jones_2022}, the WF coupling~\citep{Wouthuysen_1952, Field_1958}, Lyman-Werner feedback~\citep{Fialkov_2013}, baryon-dark matter relative velocity feedback~\citep{Fialkov_2012}, X-ray heating and ionization~\citep{Fialkov_2014},  Ly\,$\alpha$ heating~\citep{Reis_2021}, CMB heating~\citep{Venumadhav_2018, Fialkov_2019}, Ly\,$\alpha$ multiple scattering~\citep{Reis_2021}, photoheating feedback~\citep{Cohen_2016}, reionization~\citep{Fialkov_2014}, the suppression of star formation efficiencies in low mass halos~\citep{Fialkov_2013}, and redshift-space distortions~\citep{Ross_2021}.
Detailed descriptions of how \textsc{21cmSPACE} implements these physical processes and a comprehensive list of its features can be found in the most recent code development papers~\citep{Reis_2022, Gessey-Jones_2023}.
Here, we limit our discussion to the aspects of the code modified as part of this study to incorporate the Pop III IMF dependence of X-ray heating via the properties of Pop~III XRBs.

In our updated version of \textsc{21cmSPACE}, we add the capability of setting separate Pop II and Pop III specific X-ray emissivities and SEDs (previously, they were assumed to be the same).
Specifically, each star-forming halo is modelled as emitting X-rays in the energy range $0.20$ to $95.65$\,keV with a luminosity $L_{\rm X}^{\rm halo}$ in proportion to its star formation rate ${\rm SFR}^{\rm halo}$  \change{\citep[e.g.,][]{Furlanetto_2006}}
\begin{equation}
\frac{L_{\rm X}^{\rm halo}}{{\rm SFR}^{\rm halo}} = \left(3\times 10^{40} f_{\rm X, j}\right)\,{\rm erg}\,{\rm s}^{-1}\,{\rm M}_{\odot}^{-1}\,{\rm yr} \quad j \in \{{\rm II, III}\},
\label{Eq:LXSFR}
\end{equation}
where the proportionality constant $f_{\rm X, j}$ is the integrated specific X-ray emissivity of the halo\change{,} which is now set separately for Pop~II and Pop III star-forming halos as $f_{\rm X, II}$ and $f_{\rm X, III}$ respectively.
We normalize the proportionality constants in Eq.~\eqref{Eq:LXSFR} to $3\times 10^{40}\,{\rm erg}\,{\rm s}^{-1}\,{\rm M}_{\odot}^{-1}\,{\rm yr}$, as this is the theoretically and observationally predicted value for Pop II star XRBs~\citep{Fragos_2013, Fragos_2013b}, i.e., in our notation these studies predict $f_{\rm X, II} = 1$\change{, though with a non-negligible uncertainty (see below)}. \change{While Eq.~\eqref{Eq:LXSFR} was originally proposed for Pop II stars, it has also been found in simulation to hold well for Pop III stars across a range of IMFs~\citep{Sartorio_2023} (the constant $f_{\rm X, III}$ being IMF dependent), motivating our usage of it for both stellar populations.
The total luminosity derived from  Eq.~\eqref{Eq:LXSFR} is then} assumed to be distributed across frequency according to the relevant SED, SED$_{\rm II}$ for Pop II star-forming halos \change{\citep[in this work we adopt the Pop II SED from][]{Fragos_2013}} and SED$_{\rm III}$ for Pop~III star-forming halos \change{(see previous section). In \textsc{21cmSPACE}, the distribution of Pop II and Pop III star-forming halos within each simulation cell is modelled statistically using an analytic star formation prescription~\citep{Magg_2022}. Hence, the X-ray emissivity of each cell is, in practice, calculated from the cell's Pop~II and Pop III star formation rate rather than those of individual halos. Finally, }we modified \textsc{21cmSPACE} to automatically assign $f_{\rm X, III}$ and SED$_{\rm III}$  according to the Pop III IMF and the data set of specific X-ray emissivities calculated in the previous section (e.g., the $f_{\rm X, III}$ values listed in Table~\ref{tab:imfs}). As a result, both Pop III X-ray and Lyman-band emissivity are now consistently modelled from the Pop III IMF in \textsc{21cmSPACE}, as the latter was introduced into the code in our previous study~\citep{Gessey-Jones_2022}.

In addition to the Pop III IMF, \textsc{21cmSPACE} takes various high-redshift astrophysics parameters as free inputs since their values remain uncertain and \change{they} are independent of the Pop III IMF. 
\change{A description of each of these parameters is given in \textit{Supplementary Materials} Table~\ref{tab:priors}, alongside the wide priors we use for each of these during our Bayesian analysis to encompass the range of current theoretical uncertainty in their values.}
In particular, note that while we derive $f_{\rm X, III}$ and SED$_{\rm III}$ from the Pop III IMF,  $f_{\rm X, II}$ and SED$_{\rm II}$ are still independent input parameters to the code. 
For our forecasts, we treat $f_{\rm X, II}$ as an unknown to be fit for alongside the Pop III IMF while fixing SED$_{\rm II}$ to a theoretically predicted spectrum~\citep{Fragos_2013}.

\change{Since $f_{\rm X, II}$ and the Pop III IMF both impact X-ray heating, varying them has a similar, but not identical, impact on the 21-cm signal.
Consequently, a change in one of these variables can be partially compensated for by a change in the other, leaving the predicted signal only marginally altered.
As a result, if, like in this work, both $f_{\rm X, II}$ and the Pop III IMF are fit simultaneously, the posteriors on the potential values for these parameters are correlated, e.g., knowledge of one conveys a degree of knowledge of the other.
If we were instead to fix $f_{\rm X, II}$ to the data generation value, we would be implicitly leveraging additional information unavailable to us for a real 21-cm measurement. This would lead to spuriously tighter constraints on the predicted Pop III IMF. It is thus essential, for the sake of the realism of this exercise, that we simultaneously fit for any parameters of \textsc{21cmSPACE}, like $f_{\rm X, II}$, whose impact on the 21-cm signal overlaps with that of the IMF.
An additional consequence of this compensation between $f_{\rm X, II}$ and the Pop III IMF is that if the Pop III IMF is fixed to an incorrect value, the inferred value of $f_{\rm X, II}$ would be biased or vice versa (see \textit{Supplementary Materials}).
Hence, in the case of actual data wherein the correct $f_{\rm X, II}$ would be unknown, fixing its value would not only lead to erroneously tight constraints but likely bias them as well.
}

In this study, we utilize the $\Delta^2(k, z)$ 21-cm power spectrum convention
\begin{equation}
    \Delta ^2(k,z) =  \frac{k^3}{2 \pi^2}  P_{\textrm{21}}(k,z),
\end{equation}
with
\begin{equation}
    \left\langle \tilde{T}_\textrm{21}\left(\textbf{k},z\right) \tilde{T}^*_\textrm{21}\left(\textbf{k}',z\right) \right\rangle = (2 \pi)^3 \delta^{\rm 3} \left(\textbf{k} - \textbf{k}'\right) P_{\textrm{21}}(k,z),
\end{equation}
where $\tilde{T}_\textrm{21}$ is the Fourier transform of the 21-cm brightness temperature field, $\delta^{\rm 3}$ is the 3D Dirac delta-function, and $\langle \rangle$ the ensemble average.

\subsection{Emulation of the 21-cm Signal}

As discussed in the main article (and in section~\ref{ssec:nested_sampling}), we analyse synthetic 21-cm signal measurements using a nested-sampling-based methodology to ensure reliable forecasts.
A nested sampling analysis typically requires millions of likelihood evaluations and, therefore, simulations of observables.
Direct use of \textsc{21cmSPACE} within such an analysis is computationally unfeasible due to its runtime on the order of hours.
Hence, following existing 21-cm data analyses~\citep[e.g.][]{Bevins_2022}, we create neural network emulators of \textsc{21cmSPACE}.
These emulators provide a six-order of magnitude speed-up, making a nested sampling analysis feasible.
For ease of quantification of emulation error and comparison to previous works, we train a separate emulator for the 21-cm global signal and 21-cm power spectrum for each of our six Pop~III IMFs (12 emulators total).

Our twelve neural network emulators of \textsc{21cmSPACE} are based on those employed in previous 21-cm data analyses~\citep{HERA_theory_22, Bevins_2022, HERA, Bevins_2023}, with \textsc{globalemu}~\citep{global_emu} used for the 21-cm global signal emulators, and the \textsc{scikit-learn}~\citep{sklearn} multi-layer perceptron used for the 21-cm power spectrum emulators.
Each global signal emulator has five hidden layers of 16 nodes, and the power spectrum emulators have four hidden layers of 100 nodes. 
Both network architectures follow a \textsc{globalemu}-like methodology of taking the redshift, and in the case of the power spectrum emulators wavenumber, as inputs to ensure the output 21-cm signals are smooth~\citep{global_emu}. 
To train the networks, we ran 10,000 \textsc{21cmSPACE} simulations for each Pop~III IMF (60,000 simulations total), randomly sampling the input astrophysical parameters from \change{their priors listed in \textit{Supplementary Material} Table~\ref{tab:priors}}.
The 10,000 simulation outputs for each IMF, covering $7 \leq z \leq 39$ and $0.085 \leq k \leq 1$\,cMpc$^{-1}$, were then split using a test-to-train ratio of $0.1$ to form the testing and training set of the global signal and power spectrum emulator for that Pop~III IMF.

So that we can compare the accuracy of our emulators to previous works~\citep{HERA_theory_22, Bevins_2023}, we utilize the same error metrics as those studies: the root-mean-square error (RMSE) for the 21-cm global signal emulators; and the root-mean-square of the error metric
\begin{equation}
    \varepsilon_{\rm power}(k, z) = \frac{\Delta^2_\textsc{21cmSPACE}(k, z)  - \Delta^2_{\rm emulator}(k, z) }{\Delta^2_\textsc{21cmSPACE}(k, z)  + 0.1{\rm\,mK}^2},
\end{equation}
for the 21-cm power spectrum emulators. 
The latter, which we shall refer to as root-mean-square modified fractional error (RMSMFE), was introduced due to the large dynamic range of the 21-cm power spectrum and to improve emulator training by down-weighting errors when the 21-cm power spectrum is too small to be measured. 
Overall, our emulator errors are comparable to \change{those of the} aforementioned previous studies, \change{with mean RMSE $< 10$\,mK for all global 21-cm signal emulators, and mean RMSMFE < $8$\% for all 21-cm power spectrum emulators.}
\change{A full breakdown of the performance metrics computed for our twelve emulators is given in Supplementary Materials Table~\ref{tab:net_performance}.}

\subsection{Synthetic measurement data generation}

We combine a predicted 21-cm signal and an appropriate noise realization to form our synthetic measurement data sets. 
The former is generated by evaluating the emulator corresponding to the chosen data IMF and observable of interest at the synthetic data parameter values listed in \change{\textit{Supplementary Materials}}.
Where possible, these values are chosen to be consistent with theoretical or observational expectations to ensure our forecasts are as realistic as possible within current uncertainties.
The latter is specific to each experiment, its configuration, and the observation time or sensitivity.

For REACH, we assume the observation band covering frequencies $49.0$ to $167.1$\,MHz at $0.1$\,MHz resolution in agreement with the mission paper~\citep{REACH}. 
Furthermore, we assume the noise on the measured signal is a white noise Gaussian with a standard deviation of $250$, $25$, or $5$\,mK for the respective sensitivity.
Combining a realization of this white noise with a 21-cm global signal at the REACH resolution gives us our synthetic REACH observations.

Similarly, for SKA-Low, we follow pre-existing noise estimates~\citep{Koopmans_2015} for $1000$\,h of observations, scaling to the $300$ and $3000$\,h equivalents, using the fact power spectrum sensitivity goes as the reciprocal of integration time. 
These noise estimates assume observations by the SKA-Low core across $50.7$ to $177.5$\,MHz in $1.0$\,MHz frequency bins and that wavenumbers are integrated into $1$\,dex bins, with UV plane coverage of the array, the array filling factor, and thermal noise all taken into account. 
While the SKA-Low noise estimates and baselines cover a wider wavenumber range, in this study, we limit our synthetic observations to $0.1 \leq k \leq 1.0$\,cMpc$^{-1}$.
The lower limit is used to simulate a foreground avoidance strategy~\citep[see][for a discussion of foreground avoidance versus foreground mitigation]{Liu_2020} and the upper limit is a practical limitation set by the resolution of the SKA-Low core. 
The resulting noise levels vary strongly with wavenumber and redshift.
Hence, we model SKA-Low noise as achromatic Gaussian noise, adding it to 21-cm power spectrum predictions at the same frequency and wavenumber resolution to form our synthetic SKA-Low data sets.

\subsection{Nested sampling and Bayesian analysis methodology}\label{ssec:nested_sampling}

We perform a nested-sampling-based Bayesian analysis~\citep{Ashton_2022} on each of our synthetic measurement data sets using \textsc{polychord}~\citep{Handley_2015a, Handley_2015b}.
We choose to use this analysis paradigm due to its natural ability to account for parameter degeneracies, and because it is the approach taken by REACH and some existing 21-cm data analysis~\citep[e.g.][]{Bevins_2023, HERA_theory_22, Pochinda_2023}. 
In a Bayesian analysis, the prior $\pi(\theta)$, our initial knowledge of the parameter values, is revised in light of some data $D$ through the application of Bayes theorem
\begin{equation}
    \mathcal{P}(\theta | D) = \frac{\mathcal{L}(D | \theta) \pi(\theta)}{\mathcal{Z}},
\end{equation}
into our updated knowledge of the parameter values, $\mathcal{P}(\theta | D)$, the posterior.
Above $\mathcal{L}(D | \theta)$ is the likelihood (e.g.\ probability) of seeing the data given a set of parameter values, and $\mathcal{Z}$ is the Bayesian evidence, a measure of the goodness-of-fit of the underlying model to the data given by
\begin{equation}~\label{eqn:evidence}
    \mathcal{Z} = \int \mathcal{L}(D | \theta) \pi(\theta) d \theta.
\end{equation}
Constraints on the parameters can be inferred from $\mathcal{P}(\theta | D)$ (e.g.\ Fig.~\ref{fig:biases}) and alternative data models compared via $\mathcal{Z}$ (e.g.\ Fig.~\ref{fig:headline_results_plot}).

For our analysis, we treat the six Pop~III IMFs as alternative data models rather than as a parameter.
Consequently, for each of our synthetic measurement data sets, we perform six nested sampling analyses each with a fixed IMF and producing a separate posterior and Bayesian evidence $\mathcal{Z}_{i}$.
This method allows us to investigate both the biases caused by the assumption of an incorrect IMF (from the output posteriors) and the prospective constraints on the Pop III IMF (from the output evidences) using the same nested sampling runs.
Formally, the marginalized posterior probability on each IMF is given by 
\begin{equation}~\label{eqn:equivalence}
    \mathcal{P}({\rm IMF} | D) = \frac{\mathcal{Z}_{\rm IMF}}{\sum_i^{\rm IMFs} \mathcal{Z}_{i}}
\end{equation}
where the sum is over all six considered IMFs, and in using this formula we are implicitly assuming a discrete uniform prior over the IMFs.

In our analysis, we thus use a discrete uniform prior over IMFs and the priors specified in \change{Supplementary Materials} Table~\ref{tab:priors} for the other uncertain astrophysical parameters.
Additionally, in analyses of experimental data, it is typical to fit for the noise level as a free parameter.
This serves to detect systematic errors, and to improve the numerical stability of nested sampling for strongly peaked likelihoods.
Hence, we shall do likewise.
For the noise on the 21-cm global signal $\sigma_{\rm gs}$, we consider a log-uniform prior between $1$ and $1000$\,mK, and for the effective observation time for the 21-cm power spectrum $t_{\rm obs}$, we consider a log-uniform prior between $1$ and $10^5$ hours of observation. 

Alongside our priors to fully specify our Bayesian analysis, we need to establish our likelihood functions. 
We use a white noise Gaussian likelihood for our REACH analysis
\begin{equation}
    \log(\mathcal{L}(D | \theta)) = \sum_{i} \left(-\frac{1}{2}\log(2 \pi \sigma_{\rm gs}^2) - \frac{1}{2} \left(\frac{T_{\rm D}(\nu_{i}) - T_{\rm 21}(\theta, \nu_{i})}{\sigma_{\rm gs}} \right)^2\right),
\end{equation}
where $\nu_{i}$ are the REACH measurement frequencies, $T_{\rm D}$ the global 21-cm signal synthetic measurement data, and $T_{\rm 21}$ the model of the global 21-cm signal.
While for our SKA-Low analysis we use an achromatic Gaussian likelihood
\begin{equation}
    \log(\mathcal{L}(D | \theta)) = \sum_{i}\sum_{j} \left(-\frac{1}{2}\log\left(2 \pi \sigma_{i,j}(t_{\rm obs})^2\right) - \frac{1}{2} \left(\frac{\Delta_{\rm D}^{2}(\nu_{i}, k_{j}) - \Delta_{\rm 21}^{2}(\theta, \nu_{i}, k_{j})}{\sigma_{i, j}(t_{\rm obs})} \right)^2\right),
\end{equation}
where $\nu_{i}$ are now the SKA-Low measurement frequencies, $k_{j}$ the SKA-Low measurement wavenumber, $\Delta^2_{\rm D}$ the 21-cm power spectrum synthetic measurement data, $\Delta^2_{\rm 21}$ the model of the 21-cm power spectrum, and $\sigma_{i,j}(t_{\rm obs})$ the expected noise at the corresponding frequency and wavenumber after $t_{\rm obs}$ of observation.
Finally, for our joint analyses, we assume the noise between the experiments is independent, so we combine the likelihood multiplicatively.


\bibliographystyle{naturemag}
\bibliography{journals,ref}

\pagebreak

\section*{End notes} 

\subsection*{Data Availability}

The Pop III star Lyman band spectra and specific X-ray emissivities used in this work are available at 
\href{https://zenodo.org/records/5553052}{\url{https://zenodo.org/records/5553052}}
and 
\href{https://zenodo.org/records/12007614}{\url{https://zenodo.org/records/12007614}}
respectively.
All nested sampling chains resulting from our constraints forecasts are available at \href{https://zenodo.org/records/11502235}{\url{https://zenodo.org/records/11502235}}. 
The intermediate data products created as part of this study are available upon reasonable request to the corresponding author \change{AF at \href{mailto:afialkov@ast.cam.ac.uk}{\url{afialkov@ast.cam.ac.uk}}}.

\subsection*{Code Availability}

The stellar evolution codes \textsc{mesa} and \textsc{binary\_c} are available at \href{https://github.com/MESAHub/mesa/}{\url{https://github.com/MESAHub/mesa/}} and \href{https://binary_c.gitlab.io/}{\url{https://binary\_c.gitlab.io/}} respectively.
\textsc{tlusty} the stellar atmosphere code is available at \href{https://www.as.arizona.edu/~hubeny/pub/}{\url{https://www.as.arizona.edu/\~hubeny/pub/}}.
\textsc{globalemu} and \textsc{scikit-learn}, the codes we used to develop our emulators, are available at \href{https://github.com/htjb/globalemu}{\url{https://github.com/htjb/globalemu}} and \href{https://scikit-learn.org/stable/install.html}{\url{https://scikit-learn.org/stable/install.html}} respectively.
The nested sampler \textsc{polychord} is available at \href{https://github.com/PolyChord/PolyChordLite}{\url{https://github.com/PolyChord/PolyChordLite}}, and the nested sampling post-processing tool \textsc{anesthetic}~\citep{anesthetic} is available at \href{https://github.com/handley-lab/anesthetic}{\url{https://github.com/handley-lab/anesthetic}}.
\textsc{21cmSPACE} \change{and all codes created as part of this study, e.g.\ plotting scripts, are available upon reasonable request to the corresponding author AF at \href{mailto:afialkov@ast.cam.ac.uk}{\url{afialkov@ast.cam.ac.uk}}}.

\change{\subsection*{Correspondence}
Correspondence and requests for materials should be addressed
to A. Fialkov at \href{mailto:afialkov@ast.cam.ac.uk}{\url{afialkov@ast.cam.ac.uk}}.}

\subsection*{Acknowledgments}

The authors would like to thank the REACH collaboration for their feedback on an earlier draft of this manuscript and useful discussions throughout the project.
We would also like to thank the BRIDGCE UK Network for their support.

TGJ, HTJB, and EdLA acknowledge the support of the Science and Technology Facilities Council (STFC) through grant numbers ST/V506606/1, ST/T505997/1, and a Rutherford Fellowship, respectively. 
HTJB would also like to acknowledge support from the Kavli Institute for Cosmology Cambridge, and the Kavli Foundation.
NSS thanks Fonds Wetenschappelijk Onderzoek (FWO) for their support via grant 1290123N.
WJH is supported by the Royal Society University Research Fellowship.  
GMM acknowledges funding support from the Spanish State Research Agency (AEI) project PID2019-107061GB-064.
RGI acknowledges the support of the Science and Technology Facilities Council through grant numbers STFC ST/R000603/1 and ST/L003910/2.
RB acknowledges the support of the Israel Science Foundation (grant No. 2359/20).
This work was performed using the Cambridge Service for Data Driven Discovery (CSD3), part of which is operated by the University of Cambridge Research Computing on behalf of the STFC DiRAC HPC Facility (www.dirac.ac.uk). The DiRAC component of CSD3 was funded by BEIS capital funding via STFC capital grants ST/P002307/1 and ST/R002452/1 and STFC operations grant ST/R00689X/1. DiRAC is part of the National e-Infrastructure.

For the purpose of open access, the author has applied a Creative Commons Attribution (CC BY) licence to any Author Accepted Manuscript version arising from this submission.

\subsection*{Author Contributions}

TGJ extended \textsc{21cmSPACE} to model the impacts of the Pop III IMF, performed the Pop III IMF constraints forecast, and led the writing of the paper. 
NSS jointly proposed the project and simulated the Pop III XRB catalogues for the different IMFs.
HTJB provided technical support for the neural network emulation and Bayesian data analysis. 
AF jointly proposed the project, supervised it, and assisted in writing the paper. 
WJH supervised the project and provided technical support for the Bayesian data analysis. 
EdLA supervised the project and provided technical support on the 21-cm signal experiments REACH and SKA-Low. 
GMM and RGI simulated the Pop~III stellar evolution tracks and provided technical advice on using \textsc{binary\_c} for simulating Pop~III XRBs.
RB provided general advice on 21-cm cosmology theory. 
All authors contributed to revising the article by providing comments and recommendations.

\subsection*{Competing Interests}

The authors declare that they have no competing interests.

\setcounter{section}{0}
\setcounter{subsection}{0}
\setcounter{figure}{0}
\setcounter{table}{0}
\renewcommand\thefigure{S.\arabic{figure}}
\renewcommand\thetable{S.\arabic{table}}
\renewcommand\thesubsection{S.\arabic{subsection}}
\renewcommand\thesection{}



\pagebreak
\section*{Supplementary Material} \label{sec:sup_mat}

\subsection{Physical origins of 21-cm signal differences between Pop III IMFs}

\begin{figure}[ht!]
    \centering
   \includegraphics{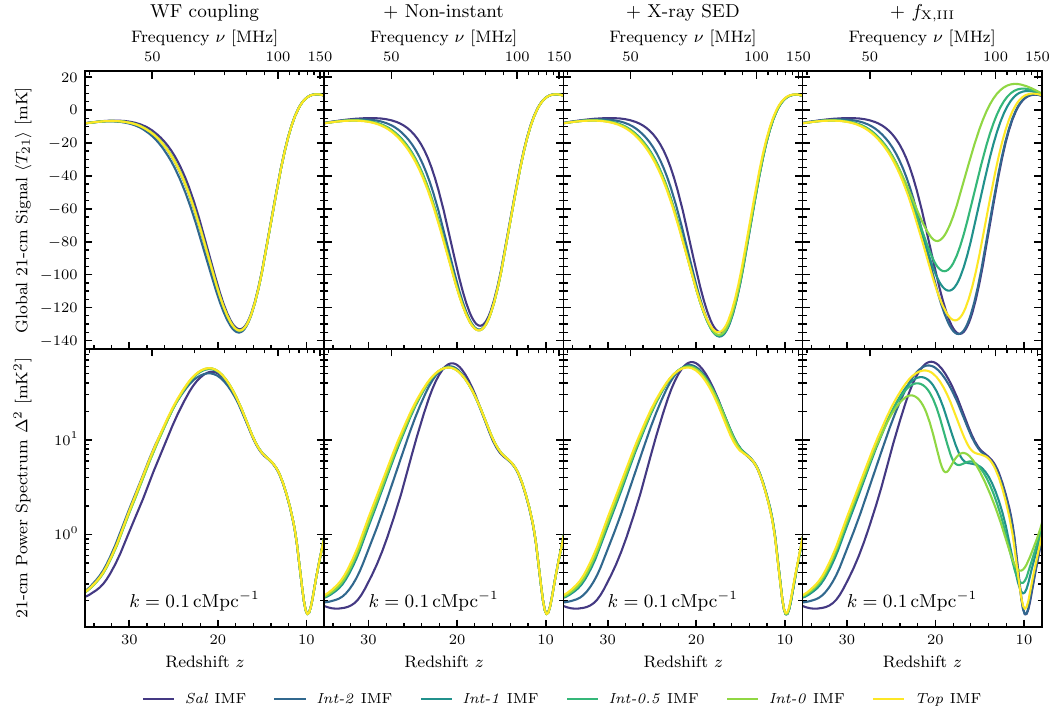}
   \caption{
   \textbf{Increasing variation of the 21-cm signal between Pop III IMFs with progressive inclusion of IMF-dependent mechanisms.}
We show the 21-cm global signal (top) and $k = 0.1$\,cMpc$^{-1}$ 21-cm power spectrum (bottom) for our six example Pop III IMFs. 
From left to right, we progressively integrate the dependence on the IMF, starting with the Lyman-band emission of the Pop III stars (WF coupling column), then the Pop III star lifetimes (Non-instant column), Pop III XRB SED shape (X-ray SED column), and Pop III XRB total X-ray emissivity ($f_{\rm X, III}$ column).
Thus, the rightmost column represents self-consistent modelling of the impacts of the Pop III IMF on the 21-cm signal and matches the signals shown in Fig.~\ref{fig:variation_with_imf}.
At $z < 23$, we find that differences in the 21-cm signal between IMFs are principally driven by Pop III XRB emissivity. 
Conversely, a combination of Lyman-band emissivity and stellar lifetime causes higher redshift variations between IMFs.
   } 
    \label{fig:variation_origin}
\end{figure}

We have shown that the 21-cm signal varies with the Pop III IMF (see Fig.~\ref{fig:variation_with_imf}). 
To understand the physical origins of these differences, we modified \textsc{21cmSPACE} so that we can select which pieces of physics are modelled self-consistently from the IMF, showing in  Fig.~\ref{fig:variation_origin} the increasing 21-cm signal deviation between IMFs as we treat additional pieces of physics as IMF-dependent.
Note, for the depicted 21-cm signals, all other astrophysical parameters (e.g., Pop II and Pop III star formation efficiencies, X-ray emissivity of Pop II stars, and the recovery time between Pop III and Pop II star formation) are fixed at the values used to generate our synthetic measurement data, as listed in Table~\ref{tab:priors}. 
When we consider only the Pop~III star Lyman-band emissivity as IMF dependant, the resultant variations in Wouthuysen-Field coupling, Ly\,$\alpha$ heating, and Lyman-Werner feedback, cause small $\Delta z < 1$ differences in the locations of the absorption trough and the high-redshift ($z\sim20$) peak of the 21-cm power spectrum. 
These differences are enhanced to $\Delta z \sim 1$ when we account for the fact that Pop III star emission is not instantaneous but instead is spread over the finite lives of the stars (as previously discussed in~\citep{Gessey-Jones_2022}).
The inclusion of this non-instantaneous emission is found to have the greatest impact on the 21-cm signal when a bottom-heavy IMF is assumed as their mass-weighted mean stellar lifetime is greater than that of top-heavy IMFs, e.g., 1320\,Myr for the \textit{Sal} IMF versus only 2.2\,Myr for the \textit{Top} IMF.
After the global 21-cm signal minimum ($z \approx 17$), we find the 21-cm signal converges between the considered IMFs due to the saturation of the Wouthuysen-Field coupling and Ly\,$\alpha$ heating being negligible compared to X-ray heating (which is not yet IMF dependent).
Hence, variations in the 21-cm signal due to the differences in Lyman-band emissivity of Pop III stars between IMFs are found to be isolated to higher redshifts.

We now consider the impact of the IMF-dependent properties of Pop III XRBs on the 21-cm signal, which are new to this study. 
The variations in SED shape (not magnitude) introduce a small $2.6$\,mK spread in the depth of the 21-cm global signals absorption trough and enhance the fixed redshift global signal differences at $z < 18$, with the largest difference seen in this redshift range being $4.8$\,mK at $z = 15$ (the $z = 15$ global signal spread was previously $1.5$\,mK without this effect). 
Similarly, enhanced divergences in the 21-cm power spectra are seen at $z < 18$ that also peak around $z \sim 15$.
These power spectra differences are found to be larger at higher wavenumbers ($3.4$ and $16.7$\,mK$^2$ at $k = 0.2$ and $1.0$\,cMpc$^{-1}$ respectively) though this is principally driven by the power spectrum increasing with $k$ at these redshifts (the corresponding fractional differences are $13$\,\% and $10$\,\%). 

Including the variation in total X-ray emissivity with the Pop III IMF leads to much greater differences between 21-cm signals, producing: a $56$\,mK spread in the depth of the 21-cm global signals absorption trough, $\Delta z \sim 3$ differences in the timing of the 21-cm global signal minimum and power spectrum cosmic dawn peak, and a secondary peak to appear in the 21-cm power spectra at $z \sim 17$ for some IMFs.
\change{This secondary power spectrum peak, often called the heating peak, occurs if heating in the early Universe is strong enough to dominate the fluctuations in the 21-cm signal~\citep{Barkana_2016}. As such, this peak is only seen for the IMFs with the most efficient X-ray emission (as seen in Fig.~\ref{fig:variation_with_imf}).}   
Furthermore, these more efficient X-ray emitting IMFs have higher and earlier emission peaks ($z \sim 12$) in the 21-cm global signal.
Hence, unlike the Lyman-band induced differences, the X-ray-introduced differences in the 21-cm signal are primarily at redshifts lower than $z = 23$ and are of greater magnitude.

Combined, the total impact of the Pop III IMF on the 21-cm signal produces large differences in both the 21-cm global signal and power spectrum from $z \sim 30$ down to $z \sim 8$.
While the exact magnitude of these differences is dependent on other uncertain high-redshift astrophysical processes, it is clear from this example that the Pop III IMF can strongly affect the observable 21-cm signal. 
Additional differences in the 21-cm signal between Pop III IMFs may also be induced at lower redshifts due to the Pop III contribution to reionization, with more top-heavy Pop III IMFs anticipated to be more efficient at emitting ionizing photons~\citep{Schaerer_2002}.
However, we do not include this effect in this work as the Pop III contribution to reionization is expected to be subdominant to that of Pop II stars, and hence, the differences in reionization history induced by varying the Pop III IMFs are anticipated to be small.

\change{
\subsection{21cmSPACE parameter priors and synthetic data values}

In Table~\ref{tab:priors}, we list the astrophysical and cosmological parameters of the \textsc{21cmSPACE} simulation code. Additionally, we give the priors used for these parameters as part of our Bayesian analysis and the values used for these parameters in the generation of our synthetic data sets. The former are intentionally broad due to the large theoretical uncertainties in many of these parameters, whereas the latter are motivated by the cited works. Details of how these priors and synthetic data sets are used within our forecasting approach, as well as the motivation for this methodology, are discussed in the main text and \textit{Methods} section.

\bgroup
\def\arraystretch{1.5}
\begin{table}[ht!]
\centering
\begin{tabular}{|c|c|c|c|c|}
\hline
Parameter & Explanation & Prior Type & Prior Range & Value used for Synthetic Data \\
\hline
\hline
$V_{\rm c}$ (km\,s$^{-1}$) & Minimum halo virial circular velocity for star formation & Log Uniform & $4.2$ to $50$ & $4.2$~\citep{Fialkov_2012, Schauer_2021, Gessey-Jones_2023}  \\
\hline 
$t_{\rm delay}$ (Myr) & Recovery time between Pop III and Pop II star formation & Log Uniform & $10$ to $100$ & $30$~\citep{Chiaki_2018,Magg_2022}   \\
\hline
$f_{*, {\rm II}}$  & Pop II star formation efficiency  & Log Uniform & $0.002$ to $0.2$ & $0.03$~\citep{Mirocha_2017, Munoz_2022, Pochinda_2023}  \\
\hline
$f_{*, {\rm III}}$  & Pop III star formation efficiency  & Log Uniform & $0.001$ to $0.1$ & $0.005$~\citep{Munoz_2022, Gessey-Jones_2022, Gurian_2023}\\
\hline
$f_{\rm X, II}$ & X-ray emission efficiency of Pop II star forming haloes & Log Uniform & $0.003$ to $300$ & $1$~\citep{Fragos_2013, Fialkov_2014}   \\
\hline
$f_{\rm rad}$  & Radio emission efficiency of high-redshift galaxies  & Log Uniform & $0.3$ to $30000$ & $1$~\citep{Gurkan_2018, Reis_2020}   \\
\hline
\hline
$\zeta$  & Effective ionizing efficiency of high-redshift galaxies  & Fixed & N/A & $15$~\citep{Furlanetto_2006, PLANCK_2018_VI}  \\
\hline
$R_{\rm max}$ (cMpc)  & Maximum mean free path of ionizing photons & Fixed & N/A & $40$~\citep{Wyithe_2004}  \\
\hline
SED$_{\rm II}$ & Spectral energy distribution of Pop II XRBs & Fixed & N/A & Fragos et al.\ 2013~\citep{Fragos_2013} \\
\hline
\end{tabular}
\caption{\change{\textbf{\textsc{21cmSPACE} physical parameters, their priors, and values used when generating our synthetic measurement data.}} 
In addition to the parameters listed here, \textsc{21cmSPACE} takes the Pop~III IMF as an input parameter, from which $f_{\rm X, III}$, SED$_{\rm III}$, and the Lyman-band emissivity of Pop III star forming haloes are derived. 
By fitting for these parameters, we can explore how assuming an incorrect IMF will bias other inferences from 21-cm data and ensure any degeneracies between these parameters and the Pop III IMF, which will weaken IMF constraints, are accounted for in our forecasts.
$\zeta$, $R_{\rm max}$ (sometimes called $R_{\rm mfp}$ in previous works), and SED$_{\rm II}$ are fixed in this study as their impacts on the 21-cm signal are small or principally at low redshifts, where the 21-cm signal is largely insensitive to the Pop III IMF.
To ensure realistic forecasts, the values used to generate the synthetic measurement data are, where available, taken to be consistent with current observations or theoretical predictions (see cited works). 
}
\label{tab:priors}
\end{table}
\egroup

}

\change{
\subsection{21-cm signal emulator performance metrics}

In \textit{Methods}, we outlined the architecture, training data, and overall performance of the neural network emulators we used in this study. 
Table~\ref{tab:net_performance} provides a detailed breakdown of the accuracy statistics of these twelve emulators.
For global 21-cm signal emulators, we utilize the RMSE error metric, and for 21-cm power spectrum emulators, RMSMFE.
We report for each emulator its mean, 68\% centile and 95\% centile error metrics over its testing set to demonstrate typical performance and the tail of the error distributions.
Previous works developing 21-cm signal emulators which included the ability to model excess radio backgrounds (as this work does), have achieved a 95\% RMSE of 20.53\,mK~\citep{Bevins_2023} for 21-cm global signal emulation and typical RMSMFE of 20\%~\citep{HERA_theory_22} for 21-cm power spectrum emulation.
While the accuracy we achieve depends on IMF, in all cases, our emulator accuracies are comparable to, or better than, these results stated in the literature.

\bgroup
\def\arraystretch{1.5}
\begin{table}[ht!]
    \centering
    \begin{tabular}{|c|>{\centering}p{2.0cm}|>{\centering}p{2.0cm}|>{\centering}p{2.0cm}|>{\centering}p{2cm}|>{\centering}p{2.5cm}|>{\centering \arraybackslash}p{2cm}|}
    \hline
         \multirow{2}{*}{IMF} & \multicolumn{3}{|c|}{21-cm global signal emulator error (mK)} & \multicolumn{3}{|c|}{21-cm power spectrum emulator error} \\
         \cline{2-7}
         & 68\% RMSE & mean RMSE & 95\% RMSE  & 68\% RMSMFE & mean RMSMFE & 95\% RMSMFE\\
         \hline 
         \hline
         \textit{Sal} & 6.97 & 8.09 & 25.71 & 0.07 & 0.07  & 0.13 \\
         \hline
         \textit{Int-2} & 4.90 & 5.95 & 15.95 & 0.07 & 0.06 & 0.11 \\
         \hline
         \textit{Int-1} & 4.92 & 6.14 & 18.04 & 0.06 & 0.06 & 0.10 \\
         \hline
         \textit{Int-0.5} & 4.75 & 5.31 & 15.16 & 0.07 & 0.07  & 0.11 \\
         \hline
         \textit{Int-0}  & 2.94 & 3.27 & 8.96 & 0.06 & 0.06 & 0.09 \\
         \hline
         \textit{Top}  & 5.09 & 5.70 & 15.62 & 0.08 & 0.07  & 0.12 \\
         \hline
    \end{tabular}
    \caption{\change{\textbf{Accuracy of neural network emulators.}}
For each of our six IMF options, we train a separate emulator of the 21-cm global signal and 21-cm power spectrum, resulting in twelve emulators in total. 
We give the 68\%, mean, and 95\% percentile values of each emulator's error statistic across its testing set. 
The error on our global signal emulators is found to be long-tailed, with a small number of the 21-cm signals in our testing sets having RMSE $> 10$\,mK.
We find these higher error global signals all display very deep minima due to having a high astrophysical radio background~\citep[$f_{\rm rad} \gg 1$,][]{Reis_2020} which enhances the magnitude of the 21-cm signal by increasing $T_{\gamma}$.
For our synthetic measurement data, we use a weak astrophysical radio background  ($f_{\rm rad} = 1$) that does not lead to any significant enhancement in the 21-cm signal magnitude~\citep[see Fig.~\ref{fig:variation_with_imf} and,][]{Sikder_2023}.
Consequently, the regions of parameter space in which the emulator is less accurate are always strongly ruled out in our analyses (confirmed via our posteriors on $f_{\rm rad}$). 
Hence, we do not anticipate this long tail in RMSE to impact our results. 
    }
    \label{tab:net_performance}
\end{table}
\egroup

}

\subsection{IMF constraints from a joint analysis of REACH and SKA-Low}

\begin{figure}[ht!]
    \centering
   \includegraphics{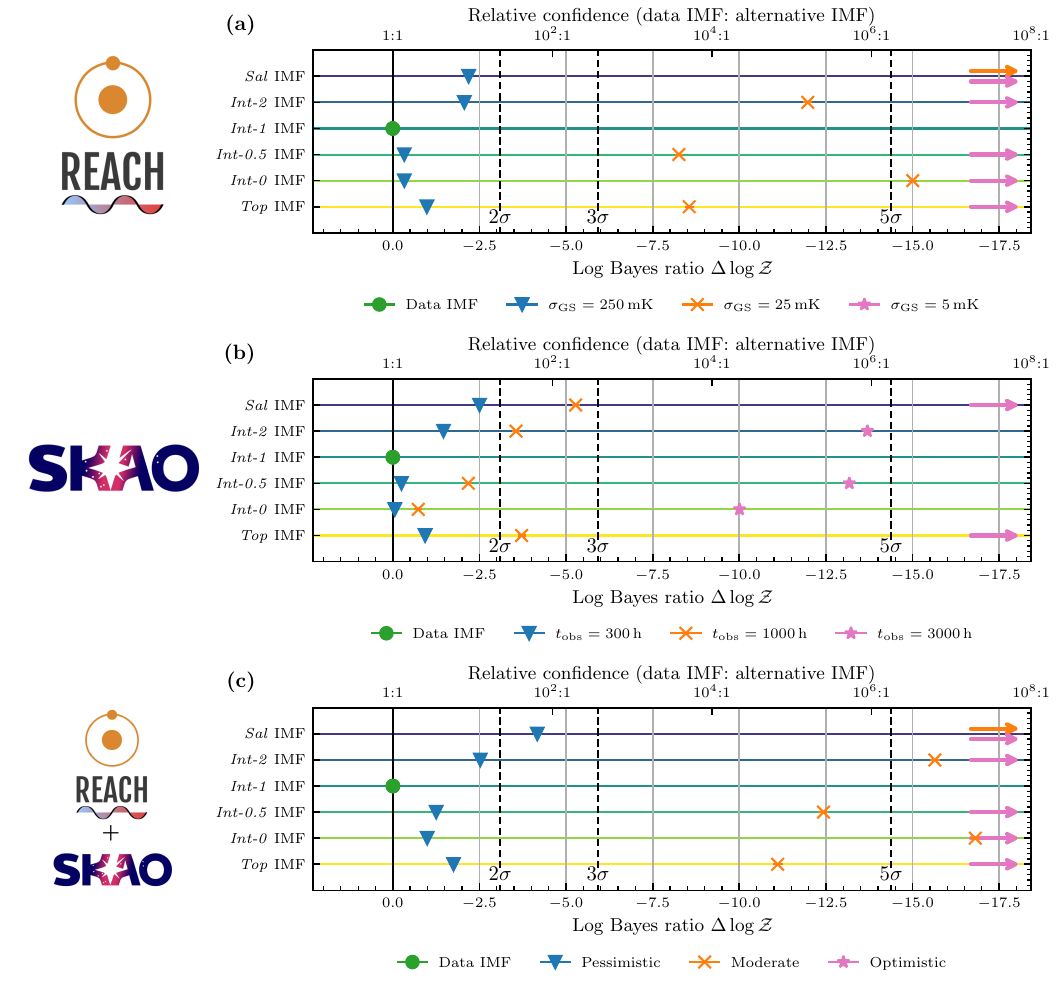}
   \caption{
   \textbf{Comparison of the prospective constraints on the Pop III IMF from individual experiments and joint analyses.} 
Each panel shows the prospective constraints on the Pop III IMF from mock 21-cm signal data analyses, expressed as the logarithm of the Bayes ratio between the fit with the data IMF (\textit{Int-1}) and the fit with an alternative IMF (top axes show the equivalent relative confidence, or betting odds, between the data IMF and the alternative IMF). 
For convenience, when the alternative IMF is rejected at $2$, $3$, and $5$\,$\sigma$ in this pairwise comparison is shown as vertical dashed lines. 
We show the REACH, SKA-Low, and joint analysis constraints in the top (\textbf{a}), middle (\textbf{b}), and bottom (\textbf{c}) rows at different sensitivities indicated by colour. 
The joint analysis improves the IMF constraints over either of the individual experiments in all cases, motivating the use of joint 21-cm data analyses in future attempts to constrain the Pop III IMF.
   } 
    \label{fig:joint_analysis_results}
\end{figure}

Alongside our forecasts for REACH and SKA-Low individually, we consider three scenarios for a joint analysis of data from these two experiments. 
In our pessimistic, moderate and optimistic scenarios, we combine $250$, $25$, and $5$\,mK sensitivity measurements by REACH with $300$, $1000$, and $3000$\,h of SKA-Low observations. 
We show the resulting forecast constraints in Fig.~\ref{fig:joint_analysis_results} alongside those from REACH and SKA-Low individually.
For this comparison, we present the prospective IMF constraints as Bayes ratios (the ratio of Bayesian evidences or, equivalently for our analysis, the ratio of posterior probabilities) between the data IMF and each alternative IMF to ensure that small changes in constraining power are visible.

The increase in the magnitude of the log Bayes ratios in the joint analyses compared to their composite individual experiments shows the disfavouring of all alternative IMFs is increased via a joint analysis. 
Hence, as discussed in the \textit{Results} section, the joint analysis provides improved constraints compared to either of the experiments individually. 
In the pessimistic case, the joint analysis gives small benefits with all improvements in log Bayes ratio less than $2$ compared to REACH at $250$\,mK on its own.
Whereas the moderate and optimistic joint analysis cases show more benefit, with all log Bayes ratios improving by more than $1.5$ (i.e., in all cases, the relative confidence in the data IMF versus an alternative IMF increases by $> 4.48$ times).
For the combinations considered here, REACH appears to be the more constraining component of the joint analysis.
However, this is likely caused by our intentionally conservative modelling of SKA-Low, with the real SKA-Low anticipated to provide significantly stronger constraints from a given amount of observation time than presented here. 
Regardless of which experiment is dominant in actuality, the improvements in constraints we see motivate the use of joint analysis of 21-cm global signal and 21-cm power spectrum data when constraining the Pop III IMF.

\subsection{Robustness of conclusions to cosmic IMF}

\begin{figure}[ht!]
    \centering
   \includegraphics{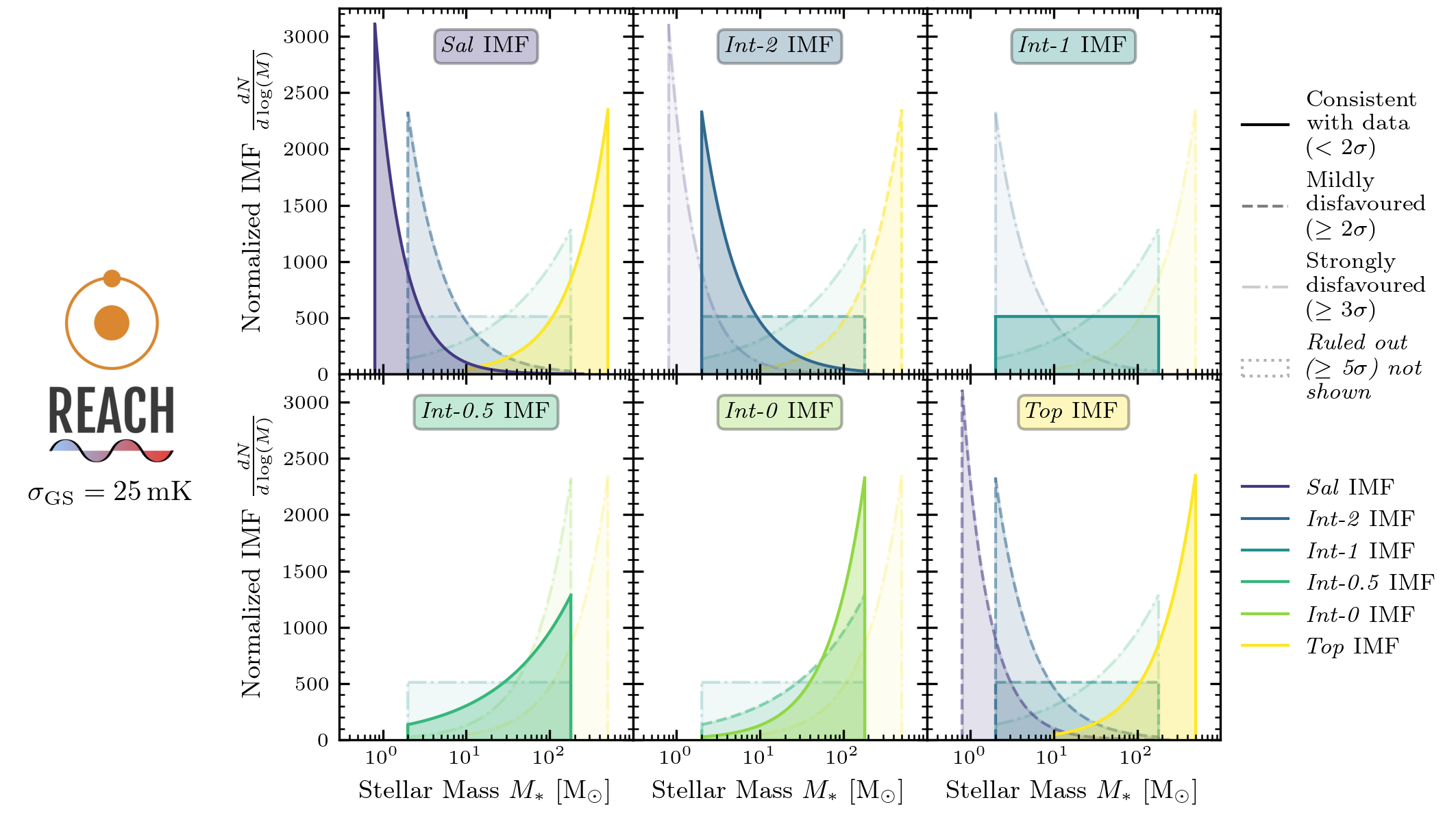}
   \caption{
   \textbf{Prospective constraints on the mass distribution of the first stars from REACH.} 
As in Fig.~\ref{fig:headline_results_plot}, each panel shows the posterior confidence in the six Pop III IMFs considered in this study, indicated via line type and opacity.
However, all constraints here are from REACH at 25\,mK sensitivity, with the different panels showing the constraints when different Pop III IMFs are used to generate the synthetic measurement data (as indicated by panel labels). 
For all six cases, we find the data IMF is the most likely, and multiple alternative IMFs are disfavoured at $> 3$\,$\sigma$.
Furthermore, all alternative IMFs are disfavoured at $> 2$\,$\sigma$, except for the \textit{Top} IMF when \textit{Sal} IMF is used to generate the data. 
Therefore, we demonstrate that a global 21-cm signal experiment, such as REACH, should be able to constrain the Pop III IMF regardless of the true cosmic Pop III IMF.
   } 
    \label{fig:reach_forecast}
\end{figure}

\begin{figure}[ht!]
    \centering
   \includegraphics{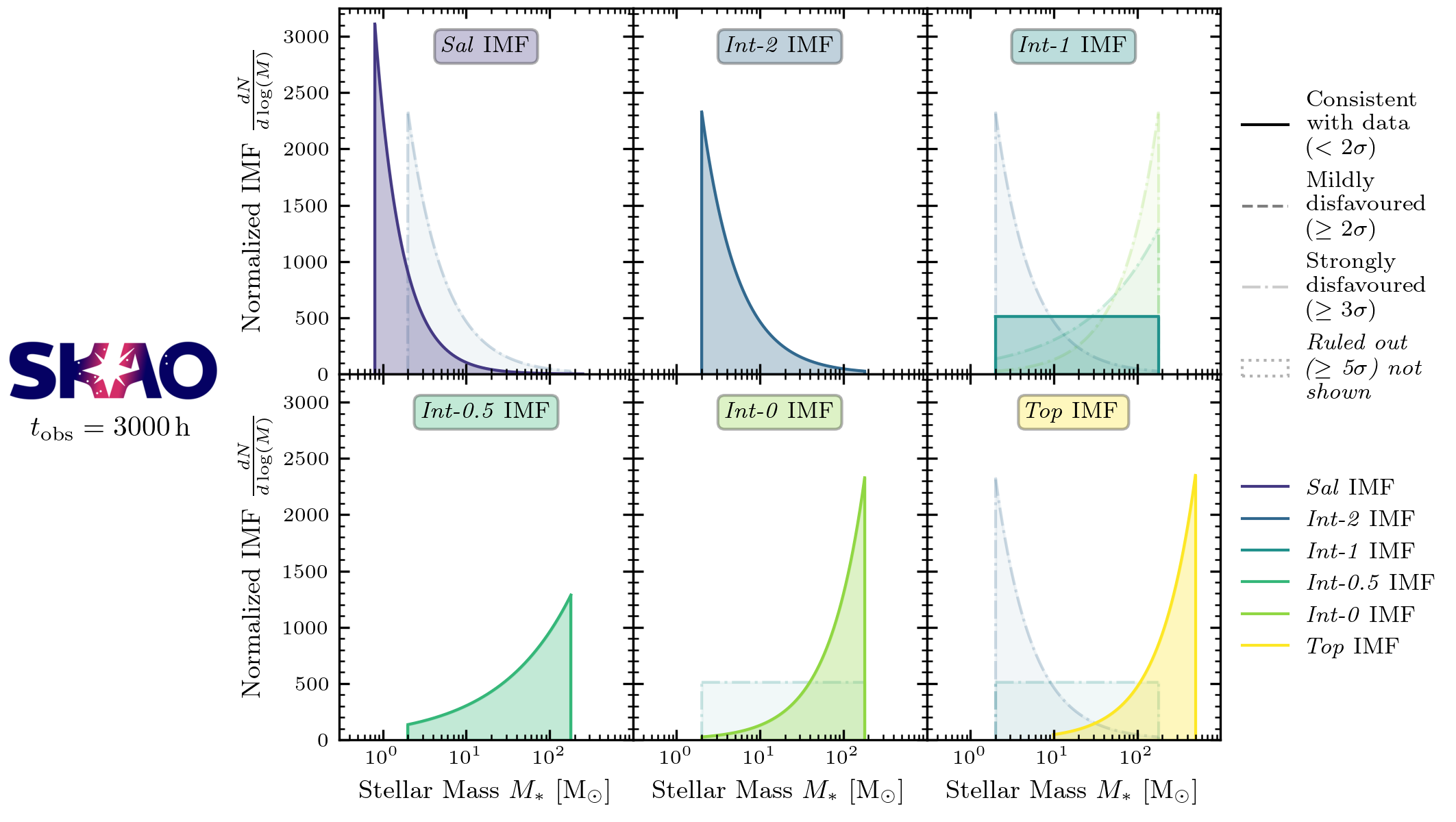}
   \caption{
   \textbf{Prospective constraints on the mass distribution of the first stars from SKA-Low.} 
As in Fig.~\ref{fig:headline_results_plot}, each panel shows the posterior confidence in the six Pop III IMFs considered in this study, indicated via line type and opacity.
However, all constraints here are from 3000\,h of foreground-avoidance observations by SKA-Low, with the different panels showing the constraints when different Pop III IMFs are used to generate the synthetic measurement data (as indicated by panel labels). 
For all six cases, we find the data IMF is the most likely, with all alternative IMFs disfavoured at $> 3$\,$\sigma$ and multiple alternative IMFs ruled out at $> 5$\,$\sigma$.
We thus find that strong constraints on the Pop III IMF are expected from SKA-Low after 3000\,h of observation, irrespective of the true cosmic Pop III IMF.
   } 
    \label{fig:ska_forecast}
\end{figure}

We previously presented the Pop III IMF constraints inferred from synthetic 21-cm signal measurement data that were generated assuming the \textit{Int-1} IMF.
However, as per the premise of this paper, the Pop III IMF of the Universe is unknown.
Hence, to show our conclusions are robust within the uncertainties surrounding the Pop III IMF, we repeat our forecasts on synthetic measurement data generated using each one of the six potential IMFs considered in this study. 
We depict the forecast IMF constraints from REACH at $25$\,mK sensitivity in Fig.~\ref{fig:reach_forecast} and from SKA-Low after $3000$\,h of observation in Fig.~\ref{fig:ska_forecast}.
For each synthetic measurement data set, the underlying 21-cm signal is simulated using the high-redshift astrophysical parameters listed in Table~\ref{tab:priors}, and we use the same priors and analysis techniques as our previous forecasting analysis (described in \textit{Methods}).

For REACH at $25$\,mK sensitivity, we find the data IMF is always correctly identified as being consistent with the synthetic measurement. 
Furthermore, at least two alternative IMFs are always disfavoured at $> 3$\,$\sigma$, and all alternative IMFs are disfavoured at $> 2$\,$\sigma$ with one exception (\textit{Top} IMF when the \textit{Sal} IMF is the data IMF). 
Our conclusion that a global 21-cm signal measurement at the expected  $25$\,mK sensitivity of REACH will be able to provide $> 3$\,$\sigma$ significance constraints on the Pop III IMF is thus robust to the actual cosmic IMF.

Similarly, we find that 3000\,h of SKA-Low observation can determine the Pop III IMF, regardless of the actual mass distribution of the first stars. 
The data IMF is always consistent with the mock measurement, and all other IMFs are disfavoured at $> 3$\,$\sigma$, with some ruled out at $> 5$\,$\sigma$.
When the synthetic measurement data was generated assuming a \textit{Int-0.5} or \textit{Int-2} IMF, the constraints are particularly strong, with all alternative IMFs ruled out at $> 5$\,$\sigma$.

\subsection{Biases in cosmic dawn parameter inference due to assuming an incorrect Pop~III IMF}

The sensitivity of an observable (e.g.\ the 21-cm signal) to a parameter (e.g.\ the Pop~III IMF) is a double-edged sword. 
It implies the potential to constrain the parameter but also that assuming an incorrect value for the parameter will likely bias estimation of other parameters from that data. 

\begin{figure}[ht!]
    \centering
   \includegraphics{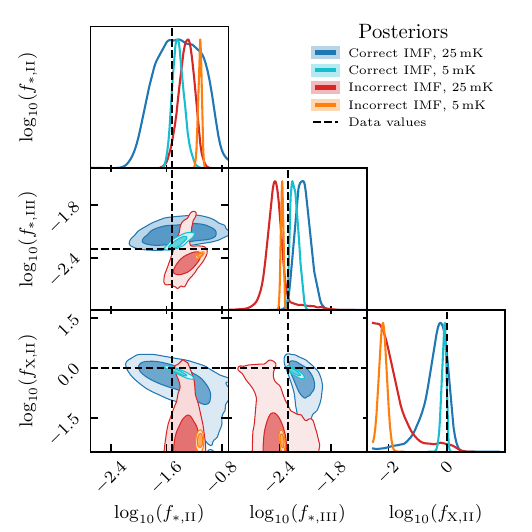}
   \caption{
   \textbf{Biased parameter constraints from 21-cm global signal observations when assuming an incorrect IMF.}
\change{Dark and light regions show $1$ and $2$\,$\sigma$ credible regions respectively.}
The 21-cm signal model used in the synthetic data for this figure assumed an \textit{Int-1} IMF, a Pop II star formation efficiency of $f_{\rm *,II} = 0.03$, a Pop III star formation efficiency of $f_{\rm *,III} = 0.005$, and a Pop II X-ray relative emissivity of $f_{\rm X} = 1$ (black dashed lines). 
When fitting this synthetic data with a model that assumes the correct IMF parameter inferences are unbiased (blue and cyan regions). 
Conversely, significant parameter biases are seen when the synthetic data is fit with a model assuming an incorrect IMF (in this case, \textit{Int-0}).
Similar biases occur in other \textsc{21cmSPACE} parameters and when the other IMFs from Table~\ref{tab:imfs} are erroneously assumed to be true. 
Note that these parameter posteriors are non-Gaussian, retroactively providing additional justification for using a nested-sampling method over Fisher forecasts~\citep{Fisher_1922} for our analysis. 
}
       \label{fig:biases}
\end{figure}

We show the values inferred for select high-redshift astrophysical parameters from our synthetic $25$\,mK and $5$\,mK sensitivity measurements of the 21-cm global signal by REACH in Fig.~\ref{fig:biases}.
When our guess of the IMF is correct, and the model has the same IMF as was used to create the synthetic data, in this case, \textit{Int-1}, we find the inferred parameter values are consistent with the values used to generate the data. 
However, when the guess is wrong, and we use an incorrect IMF to fit the data, in this case, \textit{Int-0}, we find at $25$\,mK sensitivity, the parameter values that were used to generate the data (i.e., the true values) lie outside some 2D 95\% confidence regions, with biases worsening to several sigma as sensitivity improves to $5$\,mK.
We find biased parameter constraints to be a generic consequence of assuming an incorrect IMF, occurring regardless of the data IMF, and in both REACH and SKA-Low inferences once sufficient sensitivity is reached.
Thus, to ensure reliable inferences, future 21-cm signal data analyses will need to either fit for Pop III~IMF or justify that their results are robust to the uncertainties in the Pop~III IMF. 

The underlying cause of these biases is that the model can partially compensate for an incorrect IMF choice by adjusting the values of other parameters.
However, as we saw in the \textit{Results} section, the effect of the Pop III IMF is not fully degenerate with other parameters, and this compensation is thus only partial, leaving the 21-cm signal able to distinguish between Pop III IMFs.


\end{document}